\documentclass[twocolumn,tighten,usenames,dvipsnames,twocolappendix]{aastex631}
\journalinfo{}
\makeatletter
\renewcommand{\frontmatter@title@above}{}
\makeatother
\usepackage[normalem]{ulem}
\usepackage{verbatim}
\usepackage{amssymb}
\usepackage{amsmath}
\usepackage{xspace}
\usepackage{graphicx}
\usepackage{lipsum}
\usepackage{multirow}
\usepackage{tabularx}
\usepackage{array}
\usepackage{amsmath,mathtools,mathrsfs}
\usepackage{hyperref}
\usepackage{enumitem}
\usepackage{amsthm}
\usepackage{capt-of}  
\usepackage{booktabs}

\DeclareUnicodeCharacter{2212}{-}

\DeclareMathOperator{\Arg}{Arg}

\DeclareMathOperator{\Log}{Log
}
\usepackage{leftidx}
 \usepackage{txfonts}
 \usepackage{physics}
 \usepackage{textcase}
 \usepackage{amsthm}

\definecolor{mygreen}{RGB}{150,150,10} 
\definecolor{myviolet}{RGB}{150,40,250} 

\newcommand{\s}{{\rm s}}

\newcommand{\K}{{\rm K}}

\def\mring{m-ring\xspace}

\defcitealias{M87PaperI}{M87*~Paper~I}
\defcitealias{M87PaperII}{M87*~Paper~II}
\defcitealias{M87PaperIII}{M87*~Paper~III}
\defcitealias{M87PaperIV}{M87*~Paper~IV}
\defcitealias{M87PaperV}{M87*~Paper~V}
\defcitealias{M87PaperVI}{M87*~Paper~VI}
\defcitealias{M87PaperVII}{M87*~Paper~VII}
\defcitealias{M87PaperVIII}{M87*~Paper~VIII}
\defcitealias{M87PaperIX}{M87*~Paper~IX}

\defcitealias{SgrAPaperI}{Sgr~A*~Paper~I}
\defcitealias{SgrAPaperII}{Sgr~A*~Paper~II}
\defcitealias{SgrAPaperIII}{Sgr~A*~Paper~III}
\defcitealias{SgrAPaperIV}{Sgr~A*~Paper~IV}
\defcitealias{SgrAPaperV}{Sgr~A*~Paper~V}
\defcitealias{SgrAPaperVI}{Sgr~A*~Paper~VI}
\defcitealias{SgrAPaperVII}{Sgr~A*~Paper~VII}
\defcitealias{SgrAPaperVIII}{Sgr~A*~Paper~VIII}

\theoremstyle{definition} 

\shorttitle{The nature of closure phases}
\shortauthors{Faraji \& Broderick}

\begin{document}

%

\title{
A General Response Theory for Closure Phases through Visibility Nulls and Image Structure
}
%



\author[0000-0003-1727-8992]{Shokoufe Faraji}
\affiliation{Department of Physics and Astronomy, University of Waterloo, 200 University Avenue West, Waterloo, ON, N2L 3G1, Canada}
\affiliation{Perimeter Institute for Theoretical Physics, 31 Caroline Street North, Waterloo, ON, N2L 2Y5, Canada}
\affiliation{Waterloo Centre for Astrophysics, University of Waterloo, Waterloo, ON N2L 3G1 Canada}

\author[0000-0002-3351-760X]{Avery E. Broderick}
\affiliation{Department of Physics and Astronomy, University of Waterloo, 200 University Avenue West, Waterloo, ON, N2L 3G1, Canada}
\affiliation{Perimeter Institute for Theoretical Physics, 31 Caroline Street North, Waterloo, ON, N2L 2Y5, Canada}
\affiliation{Waterloo Centre for Astrophysics, University of Waterloo, Waterloo, ON N2L 3G1 Canada}

\begin{abstract}

Closure phase is robust to station based phase errors, yet its connection to image structure is often interpreted only qualitatively and a nonzero closure phase is still interpreted mainly through qualitative symmetry arguments or source specific models. We develop, to our knowledge, the first unified response theory for an arbitrary brightness distribution and an arbitrary zero free observing triangle. Away from visibility zeros, closure phase is naturally a log-visibility observable. This structure yields an exact first order master kernel for image perturbations and a Cartesian moment hierarchy that
separates the spatial content of the source change from the sensitivity of the sampled triangle. For a fixed image, the same framework gives a local expansion in Fourier plane displacement, providing a direct description of closure phase evolution along observing tracks. We then identify the boundary of this regular regime: at a true visibility null, the log-visibility expansion fails and 
the phase behavior is controlled by the local properties of the null, including its order and winding charge, distinguishing removable phase wraps, non-null visibility minima, and nulls for which the visibility phase is multivalued. We validate these expansions using structured ring (\mring) models, showing that including successive terms converge toward the exact response. We then apply both branches of the theory to EHT observations of M87*, and demonstrate the ability to measure asymmetric image structure associated with higher image moments directly from the evolution of the closure phase. Also, the successive asymmetric moment ranks show how the closure phase of a data constrained image is constructed.
Finally, we show that the same formalism extends naturally to Phase Closure Nulling in optical and infrared interferometry. The resulting framework turns closure phase structure into a quantitative probe of faint and asymmetric source features across interferometric regimes.
\end{abstract}

 \tableofcontents


\section{Introduction}

Closure quantities are the workhorses of modern interferometry. By combining pairwise coherences so that station based calibration terms cancel, they provide observables that remain informative even when absolute phases and gains are unreliable. The best known example is the closure phase (equivalently, the bispectrum phase), formed from the triple product of complex visibilities around a baseline triangle; in the standard measurement model with station based phase errors, those phases cancel identically on any closed loop.  The analogous closure amplitude, constructed from visibility moduli on a quadrangle, cancels station based gain amplitudes.  Because closure quantities are insensitive to per-station calibration, they are widely used across radio/mm VLBI, optical long-baseline interferometry, aperture masking, and bispectrum/speckle imaging to constrain intrinsic source structure with uncertainties governed primarily by thermal (or photon) statistics when additional nonclosing systematics are negligible.

Despite this ubiquity, closure phases are often used in a largely qualitative way.  A common rule of thumb is that near zero closure phase indicates approximate $180^\circ$ point symmetry, while nonzero values indicate asymmetry.  In practice, however, closure phase time series frequently exhibit striking fork shaped discontinuities (abrupt steps of magnitude $\pi$ in the principal-value phase) especially near baseline lengths where fringe contrast becomes small. These events are routinely handled operationally (by phase unwrapping, flagging, or ad hoc amplitude thresholds), yet the literature offers limited analytic guidance on (i) how to distinguish removable $2\pi$ wraps from true singular events, (ii) what a fork implies about the underlying complex visibilities, and (iii) how to turn closure phase variations into quantitative, model-independent constraints on low order source morphology without relying on end to end imaging simulations. 

Related work across interferometry communities can be summarized as follow:
\smallskip
\noindent

\begin{enumerate}[label=\Roman*.]

\item Radio interferometry and VLBI.
Closure phase was introduced as a calibration invariant phase in \cite{Jennison1958} and has since become a cornerstone of radio aperture synthesis and VLBI practice; standard treatments of closure relations, gain invariances, and their role in imaging and self-calibration are given in \cite{ThompsonMoranSwenson2017} and in classic synthesis-imaging discussions such as \cite{Cornwell1989}.
A complementary VLBI-focused perspective, emphasizing how closure quantities enable robust imaging in the presence of station based errors, appears in broad reviews including \cite{PearsonReadhead1984}.

\item mm/submm-VLBI and the Event Horizon Telescope.
At millimeter wavelengths, rapid atmospheric phase fluctuations make station based phase calibration particularly challenging, which elevates the practical importance of closure quantities in both analysis and imaging.
Closure phase based inference and imaging play a central role in modern horizon scale VLBI, including early 1.3\,mm VLBI demonstrations \cite{Doeleman2008} and the EHT imaging campaigns on M87* and Sgr~A* \cite{EHT2019IV,EHT2022III}.
Algorithmically, several modern imaging frameworks incorporate closure phases (often together with closure amplitudes or robust likelihoods) as primary data products; see, e.g., \cite{Chael2018} and references therein.

\item Optical/IR long baseline interferometry and aperture masking.
Closure phase is equally fundamental in optical/IR interferometry, where atmospheric turbulence similarly corrupts station based phases.
Early optical demonstrations of closure phase imaging and high fidelity model constraints include seminal aperture-synthesis efforts \cite{Baldwin1986,Haniff1987}, and the broader methodological landscape is reviewed in \cite{Monnier2003} (see also \cite{BaldwinHaniff2002}).
Sparse aperture (masking) interferometry extends these ideas to filled-aperture telescopes and has become a standard route to robust phase information in high angular resolution imaging \cite{2000SPIE.4006..491T}.

\item Speckle interferometry, bispectrum phase retrieval, and speckle masking. The closure phase is the phase of the bispectrum (triple product), and closely related bispectrum/triple-correlation techniques were developed in the speckle interferometry literature for phase retrieval through atmospheric turbulence \cite{Labeyrie1970,Weigelt1977,LohmannWeigeltWirnitzer1983,Wirnitzer1985}.
This parallel development underscores that closure/bispectrum phases provide a broadly applicable route to calibration-robust Fourier-phase information across interferometric modalities.

\item Geometric and invariant viewpoints.
Recent work has emphasized geometric and graph/gauge theoretic perspectives on closure quantities and their invariances (beyond traditional algebraic derivations), providing complementary conceptual frameworks for closure relations and their interpretation \cite{Thyagarajan2022}.
\end{enumerate}
The principal result of this paper is a general quantitative interpretation of closure phase structure without assuming a binary, ring, or other source specific model. A single visibility space response law is pulled back in two complementary ways: to arbitrary changes in the image at fixed baselines, and to motion of a fixed image through the Fourier plane. The topology of visibility zeros completes this construction by identifying the singular boundary of the regular response. Therefore, the aim of this paper is to provide a common framework for interpreting the structures already present in interferometric data. We separate changes produced by the source from those produced by motion of the observing triangle, and we treat the regular response away from visibility zeros together with the distinct behavior at a true null. This organization turns closure phase into an operational diagnostic: it identifies what kind of phase feature has occurred, which source structures can produce it, and when a local approximation can be trusted. There are model-independent moment relations have previously been developed in the marginally resolved limit e.g., \cite{2003A&A...400..795L}, where closure phase first appears at third order in baseline and probes image asymmetry. The present construction is not restricted to an expansion about zero baseline or to a marginally resolved source. We retain the sampled reference visibilities exactly and develop the response about an arbitrary zero free triangle, including strongly resolved and near null regimes. This produces an exact first order functional kernel for arbitrary image perturbations, a full Cartesian moment hierarchy, and the complementary response of a fixed image along a moving observing triangle.

\medskip

In this work, the general sky brightness distribution on the plane of the sky, denoted by $I(\mathbf{x})$, where $\mathbf{x}$ is the angular position. No assumption is made here about the source such as a binary, a collection of point components, or any other special parametric form. The topology discussed in this paper depends only on the behavior of the complex visibility as a function of baseline, and especially on the structure of its zeros.
Thus the arguments apply to arbitrary brightness distributions, including extended and asymmetric images.

In this paper, the baseline formed by telescopes $i$ and $j$ at time $t$ is represented by the dimensionless projected baseline vector
\begin{equation}
\boldsymbol{u}_{ij}(t)=\frac{\boldsymbol{B}_{ij}(t)}{\lambda},
\end{equation}
where $\boldsymbol{B}_{ij}(t)$ is the geometric baseline vector on the plane perpendicular to the
line of sight, and $\lambda$ is the
observing wavelength. The vector $\boldsymbol{u}_{ij}(t)$ is often called the
spatial frequency, or the $uv$-coordinate, because it is Fourier conjugate to
angular position on the sky. However, the time dependence simply reflects the fact that, as the Earth rotates, a fixed pair of telescopes samples different
points in the Fourier plane. The corresponding complex visibility is the Fourier transform of the sky brightness evaluated at this spatial frequency,
\begin{equation}\label{eq:visibility-fixed-image}
V_I(\boldsymbol u)=\int_{\mathbb R^2}
I(\boldsymbol x)\, e^{2\pi i\,\boldsymbol u\cdot\boldsymbol x}\,
d^2\boldsymbol x.
\end{equation}
This notation will be used throughout the paper. Only when an explicit example is needed later, we may specialize to a simple source model whose visibility zeros can be written in closed form. Because the framework applies across radio, optical/infrared, speckle, and aperture masking interferometry, \autoref{tab:result-scope} provides a compact cross domain guide to the terminology, assumptions, and scope of the principal results.

This paper is organized as follows. \autoref{sec:topolgy-thm} develops the relevant topological properties of the complex visibility, while \autoref{sec:functionalkernel} presents the main analytic framework and response results.
The formalism is illustrated through practical applications in \autoref{sec:example1}, and the main conclusions are summarized in \autoref{sec:summary}.

\section{Topology of Closure Phase Singularities}\label{sec:topolgy-thm}

In this section, the word topology has an operational meaning and we are not referring to the topology of the image on the sky or to any global topology of the source. It means a property of the complex visibility that is unchanged under continuous deformations.

\subsection{Visibility zeros}\label{sec:visibility-zeros}

The first objects of interest are the zeros of the complex visibility. With the
notation of \autoref{eq:visibility-fixed-image}, a visibility zero is a point in the spatial frequency
plane at which
\begin{equation}
V(\boldsymbol{u})=0 .
\end{equation}
Equivalently, the visibility amplitude vanishes at that point. The interferometric phase is then not defined, since the phase of a complex number can only be assigned when the
complex number is nonzero.

For a general brightness distribution, the positions of these zeros depend on the full
structure of the image. This is not a
limitation for the present discussion. What matters for this study is only that the
visibility is a complex function on the spatial frequency plane, and that its zeros are the points around which the visibility phase can wind. When a closed path in the
$\boldsymbol{u}$-plane encircles such a zero, the phase of the visibility need not return
trivially; it can acquire a winding by an integer multiple of $2\pi$. This integer
winding is the local topological charge of the visibility zero.

In a generic image, the real and imaginary parts of $V(\boldsymbol{u})$ both have to
vanish at the same point. Thus the zeros are usually isolated points in the
two-dimensional $\boldsymbol{u}$-plane. Additional symmetries can change this behavior.
For example, highly symmetric images may produce null curves or rings rather than isolated
zeros. Such cases are special, and can usually be regarded as degeneracies that split into
isolated zeros once the symmetry is slightly broken. The discussion below is
therefore naturally phrased in terms of isolated visibility zeros, while allowing special
symmetric limits as limiting cases.

An interferometric observation does not sample the whole $\boldsymbol{u}$-plane at once.
Each baseline traces a track $\boldsymbol{u}_{ij}(t)$ as the Earth rotates, and the
measured visibility on that baseline is the value of $V$ along this track. A visibility
zero is encountered when the baseline track passes through one of the zeros of
$V(\boldsymbol{u})$. Near such a point the visibility amplitude becomes small and the
baseline phase becomes undefined. At the zero itself the baseline phase is singular.

This is the origin of the closure phase singularities studied in this section. A closure phase is built from the phases of three visibilities on a closed triangle of baselines $\triangle=(i,j,k)$
\begin{equation}
B_\triangle(t)=V_{ij}(t)\,V_{jk}(t)\,V_{ki}(t),
\label{eq:bispectrum-general}
\end{equation}
and the closure phase is the pha se of this complex number,
\begin{equation}\label{eq:CP_initial}
\phi_\triangle(t)=\Arg B_\triangle(t).
\end{equation}
This definition also makes clear where the closure phase can become singular. If one of these visibilities passes through a zero, then one of the three phases entering
the closure phase becomes undefined
\begin{equation}
B_\triangle(t)=0
\quad\Longleftrightarrow\quad
V_{ij}(t)=0\ \text{or}\
V_{jk}(t)=0\ \text{or}\
V_{ki}(t)=0 .
\label{eq:bispectrum-zero-set}
\end{equation}
The behavior of the closure phase near such an event is controlled not by the detailed source model, but by the winding of the complex
visibility around the zero, in the following sense. This is also worth mentioning that some apparent phase features can be changed or removed by the choices like the image phase center are intrinsic to local structure of the visibility field near the null. We return to this discussion in the next subsection. 


\subsection{True visibility nulls and pseudo nulls and some examples}

We will use the word null in a specific sense. A true visibility null is a point in the Fourier plane where the complex visibility itself vanishes,
\begin{equation}
V(\boldsymbol{u})=0 .
\end{equation}
Equivalently, the visibility amplitude $|V(\boldsymbol{u})|$ is zero. At such a point the visibility does not have a well-defined phase. These are the nulls that can produce closure phase singularities.

This should be distinguished from a zero of only one plotted component of the visibility. For example, the real part of $V(\boldsymbol{u})$ may vanish while its imaginary part remains nonzero. In that case, the complex visibility has not vanished, its phase is still defined,
and there is no true null. We will refer to such zero crossings as pseudo nulls. They can be useful in visualizing the visibility and some practical closed forms, but they are not the singular points relevant for the closure phase.

Indeed, the distinction is especially important because changing the phase center of the image can move these pseudo nulls around. If the image is shifted on the sky by an angular vector $\mathbf{x}_0$, the visibility is multiplied by a phase. This factor has unit magnitude, i.e., it can rotate the complex visibility and change where the real or imaginary part crosses zero, but it cannot change $|V|$. Thus the  choice of phase center cannot create or remove a true visibility null, while it can only create or remove
pseudo nulls in a chosen projection of the complex visibility.

A simple example is an equal flux binary. When the two fluxes are equal, $I_1=I_2=I$, the visibility has true nulls on the lines , i.e., $\boldsymbol{u}\cdot\boldsymbol{\Delta}=n+1/2$, for $n\in\mathbb{Z}$. At these baselines the two components contribute equally with opposite phase, so the complex visibility itself cancels.
Now, if the two fluxes are not equal, this exact cancellation is lost. For example, if the brighter component is placed at the origin and the dimmer component is displaced by $\boldsymbol{\Delta}$, (i.e, $I_0>I_1$), then the visibility can be written as
\begin{equation}
V(\boldsymbol{u})=I_0 + I_1 e^{2\pi i\,\boldsymbol{u}\cdot\boldsymbol{\Delta}},
\end{equation}
As the baseline changes, the second term traces a circle of radius $I_1$ around the constant offset $I_0$ in the complex plane. Since the radius is smaller than the offset, the circle never reaches the origin. The visibility amplitude remains nonzero. If one chooses a different phase center, the real part of this same visibility may cross zero, but those zero crossings are pseudo nulls, not true visibility nulls.

The same reasoning applies more generally. Suppose a source contains a dominant central component of flux $I_0$, together with several weaker components whose total flux is less than $I_0$. Then the oscillatory terms from the weaker components cannot cancel the constant contribution from the central one. In particular, a collinear three component source with a central flux larger than the combined flux of the two outer components has structure, and its visibility can oscillate with baseline, but again it need not have any true nulls.

By contrast, rings and disks provide standard examples where true nulls happen naturally. For a thin circular ring of angular radius $R$ and total flux $F$ which is appropriate example for black hole image, the visibility depends only on $\rho=|\boldsymbol{u}|$ and is
\begin{equation}
V_{\rm ring}(\rho)\propto\,J_0(2\pi R\rho),
\end{equation}
where $J_0$ is the Bessel function of the first kind. The true nulls are placed at $2\pi R\rho = j_{0,n}$,
where $j_{0,n}$ is the $n$-th zero of $J_0$. Therefore, a centered thin ring produces circular null curves in the Fourier plane.

A uniform circular disk gives a similar example which is appropriate model for stars. For a disk of angular radius $R$ and total flux $F$,
\begin{equation}
V_{\rm disk}(\rho) \propto \,\frac{2J_1(2\pi R\rho)}{2\pi R\rho},
\end{equation}
with the value at $\rho=0$ defined by continuity. Its nulls happen when $2\pi R\rho = j_{1,n}$, where $j_{1,n}$ is the $n$-th zero of $J_1$. These are true nulls of the complex visibility amplitude, not only zeros of a plotted real component. These nulls are not tied to any binary structure; they are
a direct consequence of destructive interference between emission from different parts of the ring.

\medskip

These examples show that visibility zeros happen in very simple and physically familiar brightness distributions. At the same time, the existence of a null is not always guarantee; smooth or strongly dominated images can avoid them. A single Gaussian source is the simplest counterexample. A centered circular Gaussian has a Gaussian visibility,
\begin{equation}
V_{\rm gauss}(\rho)\propto\exp\left[-2\pi^2\sigma^2\rho^2\right],
\end{equation}
and this never vanishes at any finite baseline. If the Gaussian is shifted away from the image center, the visibility real part will contain a cosine factor and may have many zero crossings. But the visibility amplitude will be still the same Gaussian envelope, so the complex visibility has no true null. These are pseudo nulls introduced by looking at a particular phase rotated component.

Therefore, the true visibility nulls are common in many
structured images, especially when different regions of the image contribute comparable Fourier amplitudes that can cancel. For a generic asymmetric image these degeneracies are usually broken, and the zeros, when present, become isolated points. It is the phase winding around such zeros that controls the closure phase singularities discussed in the next subsections. 

The nulls relevant for the rest of this paper are the true nulls of the complex visibility amplitude.

\subsection{Closure phase forks and phase wrapping}
\label{sec:closure phase-forks}

We now describe the closure phase signature of a true visibility null. The important point is that a discontinuity in a plotted closure phase is not necessarily evidence for a bispectrum /zero. Since closure phase is an angle, its value is always given modulo $2\pi$. Therefore, a smooth phase evolution can also appear as jump when mapped into a fixed plotting range like $(-\pi, \pi]$. Such a jump has no invariant meaning and it can be removed by choosing a continuous branch of the phase. This is the usual branch cut effect.

It is useful to state this wrapping more precisely with winding number language. On any time interval that bispectrum does not vanish, it can be written in amplitude phase form
\begin{equation}
 B_\triangle(t)
=|B_\triangle(t)|\,\,e^{i\widetilde{\phi}_\triangle(t)},
\qquad |B_\triangle(t)|>0,
\end{equation}
here $\widetilde{\phi}_\triangle(t)$ is chosen to vary continuously along the interval. The closure phase that is plotted $\phi_\triangle(t)$ is only its value modulo $2\pi$. Thus
\begin{equation}\label{eq:wrapp-unwrapped}
\widetilde{\phi}_\triangle(t)
=\phi_\triangle(t)+2\pi n_\triangle(t),
\qquad n_\triangle(t)\in\mathbb Z,   
\end{equation}
where tilde means the unwrapped, continuously followed closure phase on an interval where the bispectrum is nonzero. When the plotted phase jumps by $2\pi$ while $|B_{\triangle}|$ remains nonzero, $n_\triangle$ changes so that the unwrapped phase remains continuous. If over an interval $[t_1,t_2]$ the bispectrum returns to the same direction in the complex plane 
\begin{equation}
  \frac{B_\triangle(t_2)}{|B_\triangle(t_2)|}
=\frac{B_\triangle(t_1)}{|B_\triangle(t_1)|}.  
\end{equation}
The the net unwrapped phase advance is an integer multiple of $2\pi$
\begin{equation}
N_\triangle
=\frac{\widetilde{\phi}_\triangle(t_2)-\widetilde{\phi}_\triangle(t_1)}{2\pi}
\in\mathbb Z.
\end{equation}
This integer is winding number of the bispectrum direction over that closed evolution. It counts the number of full turns made by the bispectrum around the origin and describes accumulated phase away from singularities. For a general open interval there is no integer winding invariant since the unwrapped phase depends on the chosen starting value. More importantly this construction requires $|B_{\triangle}|\neq0$.

However, the situation changes when the bispectrum itself vanishes. In this case, the $\pm\pi$ jump at a true zero cannot be removed in this way, because the closure phase is undefined at the crossing itself. On the two sides of the zero the phase can be followed continuously, but the two sides are separated by the singular point. We will call such a nonremovable event a closure phase fork. Thus, a fork can happen only when one the visibility factors in the bispectrum has a true zero. The generic local form is as follows. Suppose that the null happens on baseline $(i,j)$ at time $t_\ast$, while the other two baselines in the triangle remain nonzero. For a simple crossing ($\dot V_{ij}(t_\ast)\neq 0$) near $t_\ast$, the vanishing visibility is well approximated by its
linear behavior,
\begin{equation}
    V_{ij}(t)=
    \dot V_{ij}(t_\ast)(t-t_\ast)+ \mathcal O\left((t-t_\ast)^2\right)
\end{equation}
The other two factors, $V_{jk}(t_\ast)$ and $V_{ki}(t_\ast)$, are nonzero complex numbers. Locally they only multiply the result by a fixed amplitude and phase. Hence the bispectrum has the same
linear zero,
\begin{equation}\label{eq:fork_bis}
B_\triangle(t)=
\dot V_{ij}(t_\ast) V_{jk}(t_\ast)
V_{ki}(t_\ast)(t-t_\ast)
+\mathcal O\left((t-t_\ast)^2\right).
\end{equation}
This equation has a simple interpretation. As $t$ passes through $t_\ast$, the real factor $t-t_\ast$ changes sign. Multiplying a complex number by a negative real number rotates its phase by $\pi$. Therefore the bispectrum approaches the origin in the
complex plane from one direction and leaves in the opposite direction. The two limiting
closure phases differ by $\pi$. Note that a baseline track that passes through the zero samples only a one dimensional cut of this phase field, and this cut produces the half-turn observed as closure phase fork and gives the nonremovable $\pm\pi$ closure phase jump
associated with this true fork
\begin{equation}\label{eq:time-track-fork}
\lim_{t\to t_\ast^+}\phi_\triangle(t)
-\lim_{t\to t_\ast^-}\phi_\triangle(t)=\pm\pi
\end{equation}
where $\phi$ is given by \autoref{eq:CP_initial}. Of course, the pseudo nulls do not produce this effect, because they leave the visibility amplitude nonzero.

It is worth mentioning that if the visibility on baseline $(a,b)$ has a true null, then the same vanishing factor appears in the bispectrum of every triangle that contains $(a,b)$. Thus, in an array with $N$ stations, a single baseline null is copied into $N-2$ closure phases. The diagnostic is therefore not an isolated jump in one plotted closure phase, but a correlated array pattern: the affected bispectra lose amplitude, the corresponding closure phases show nonremovable $\pi$-type forks, and the forked triangles share a common baseline edge. This also explains why the sign of the plotted jump is not the primary observable. Different triangles containing the same nulling baseline can show different signs, because the remaining visibility factors have different phases.




\subsection{Topological protection of visibility nulls}
\label{sec:topological-protection}


\begin{figure*}
    \centering
    \includegraphics[width=\textwidth]{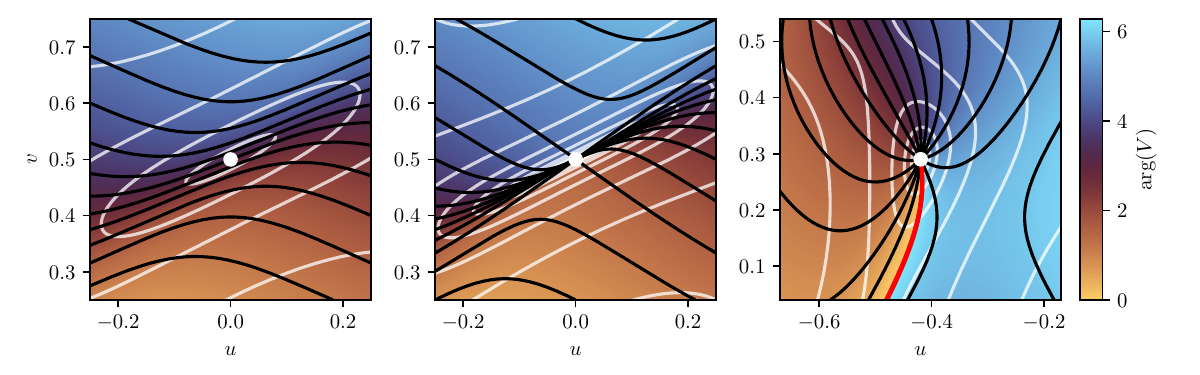}
    \caption{Visibility phase contours (black) and amplitude contours (white) for a trinary exhibiting a non-null visibility minimum (left), a visibility null with zero winding number (center), and a visibility null with a phase fork, i.e., a winding number of $-1$ (right).  The location of the minimum visibility amplitude is shown by the white circle and centered in each panel.  A branch cut appears in the right panel and is indicated by the redline. }
    \label{fig:forks}
\end{figure*}

As mentioned in the beginning of this section, the word topology here means a property of the complex visibility that is unchanged under continuous deformations. For instance, the detailed position of a visibility zero in the Fourier plan may move if the image is slightly changed. But an isolated zero with nonzero phase circling around it cannot simply disappear under such a change, for example, it can be annihilated with another zero carrying the opposite circulation. We know make this statement precise for an isolated true visibility null as follows.

The winding number discussed in \autoref{sec:closure phase-forks} was associated with following a phase along an observing time track. The topological question for a visibility null is different. It is a local question in the two dimensional Fourier plane as if the complex visibility vanishes at $\boldsymbol{u}_0$, can this zero be removed by a small change of convention or by a small perturbation of the image?

To answer this one should not look at the phase at the null itself, because the phase is undefined there. Instead, take a small closed curve $\gamma$ around $\boldsymbol{u}_0$ chosen in a way that no other visibility zero lies on or inside $\gamma$. The phase of $V$ is not defined at $\boldsymbol{u}_0$, but it is defined on $\gamma$. We choose the orientation of $\gamma$ to be counterclockwise and parametrize it by $s\in[0,1]$, with $\gamma(1)=\gamma(0)$. Since $V$ is nonzero on this curve, it can be written along the curve in amplitude phase form,
\begin{equation}
V(\gamma(s))=|V(\gamma(s)) |\,e^{i\alpha(s)} ,
\end{equation}
where $\alpha(s)$ is chosen continuously as one goes around the loop. Since the curve is closed, the visibility returns to the same complex value after one circuit. The phase can therefore only change by an integer multiple of $2\pi$
\begin{equation}
\alpha(1)-\alpha(0)=2\pi q.
\end{equation}
We call this integer
\begin{equation}
q(\boldsymbol{u}_0)=\frac{\alpha(1)-\alpha(0)}{2\pi}
\end{equation}
the local charge of the visibility zero. Equivalently,
\begin{equation}\label{eq:charge-integral}
q(\boldsymbol{u}_0)=\frac{1}{2\pi}\oint_\gamma d\,\arg V.
\end{equation}
This charge is a property of the phase pattern around zero and not of one particular baseline track. The loop $\gamma$ lives in two dimensional Fourier plane while an observation samples one dimentional track through this plane. Note that the charge is not the same as the $\pi$ jump seen along a time track (\autoref{eq:time-track-fork}). The charge is mesured by circling the zero in the Fourier plane. This means the charge characterizes the visibility zero itself while the closure phase fork is what is seen when a baseline track crosses such a zero.  



If the total charge inside $\gamma$ is nonzero, the visibility cannot be made nonzero everywhere inside the loop by a small smooth change of the image, as long as the visibility remains nonzero on the loop, the total charge inside the loop is conserved. A charged zero can disappear
from the region only by crossing the loop or by meeting other zeros whose total charge cancels it. We will explain this in more detail in the next subsection.


There is a precise distinction between a true nul and a pseudo null. A pseudo null is a zero of a chosen component or phase rotated projection of the complex visibility, for example, the real part of $V$ may vanish, while $|V|$ remains nonzero. It has no singular phase field around it and no local charge. A charged true null is different: the complex visibility itself vanishes, and the surrounding phase field has nonzero circulation.

A true zero with zero net charge is also possible. Such a point is not a pseudo null, because the complex visibility may still vanish there. However, it is not protected in the same way. With zero net charge there is no topological obstruction to removing the zero by a small physical perturbation. These cases are typically accidental degeneracies, or unresolved combinations of nearby zeros with opposite charges.



\subsection{Null order and winding number}\label{sec:order-winding}

We now separate two notions: the order of a visibility null and the winding number carried by that null. The order tells us how the amplitude of $V$ vanishes near the null.The winding number tells us how the phase of $V$ turns around it. Although these two are related, they are not the same.

Let $V(\boldsymbol u)$ be the complex visibility, with
$\boldsymbol u=(u,v)$, that has an isolated null at
\begin{equation}
\boldsymbol u_\ast=(u_\ast,v_\ast), \qquad
V(\boldsymbol u_\ast)=0 .
\end{equation}
and introduce local coordinates centered on the null as
\begin{equation}
\xi=u-u_\ast, \qquad \eta=v-v_\ast.
\end{equation}
We assume here that $V$ is smooth enough near the null to have the local expansion\footnote{This is automatic for finite point source models and for sufficiently regular compact images.}. Taylor expansion of $V$ around the null begins at some positive order
\begin{equation}\label{eq:V-homo-expansion}
V=P_m(\xi,\eta)+ P_{m+1}(\xi,\eta)+P_{m+2}(\xi,\eta)+\cdots.
\end{equation}
Here $P_m$ is the first nonzero term. It is a homogeneous polynomial of degree $m$
\begin{equation}
P_m(\lambda\xi,\lambda\eta) = \lambda^m P_m(\xi,\eta),
\end{equation}
equivalently,
\begin{equation}
P_m(\boldsymbol \xi)=\frac{1}{m!}\left.
\nabla_{\boldsymbol u}^{\,m}V \right|_{\boldsymbol u=\boldsymbol u_\ast}:\boldsymbol \xi^{\otimes m}=a_{i_1\cdots i_m}\,\xi^{i_1}\cdots \xi^{i_m},
\label{eq:Pm-tensor-form}
\end{equation}
where $a_{i_1\cdots i_m}$ is a complex symmetric rank-$m$ tensor. The integer $m$ is the order of the null. A simple null has $m=1$, i.e., quadratic null has $m=2$, and so on.

The winding number is defined differently. We take a small counterclockwise circle around the null,
\begin{equation}
\xi=\epsilon\cos\theta, \qquad
\eta=\epsilon\sin\theta, \qquad
0\leq\theta\leq2\pi.
\end{equation}
If $\epsilon$ is small enough that the circle contains no other nulls and $V\neq0$ on the circle, then the local charge is given by \autoref{eq:charge-integral}, which $q$ counts how many times the complex number $V$ winds around the origin as the
point $\boldsymbol u$ goes once around the null.

In fact, the leading term $P_m$ determines the winding if it is nonzero in every direction around the null
\begin{equation}
P_m(\cos\theta,\sin\theta)\neq0, \quad \text{ for all } \theta.
\end{equation}
Then $P_m$ alone controls the local winding\footnote{If the leading term $P_m$ vanishes for some direction in the unit circle, then the leading term alone is not enough to determine the winding. In that case one should compute the winding from the full visibility $V$ on a sufficiently small loop using \autoref{eq:charge-integral}.}. In that case, on the small circle
\begin{equation}
V\simeq \epsilon^m\left[P_m\,(\cos\theta,\sin\theta)\right],
\end{equation}
here $\epsilon^m$ is real and positive, so it does not affect the phase. For sufficiently small $\epsilon$, the winding of $V$ is therefore the winding of the curve $P_m(\cos\theta,\sin\theta)$ in the complex plane. Therefore
\begin{equation}
q(\boldsymbol u_\ast)=\frac{1}{2\pi i}\int_0^{2\pi}
\frac{\partial_\theta P_m(\cos\theta,\sin\theta)}{
P_m(\cos\theta,\sin\theta)}\,d\theta,
\label{eq:winding-from-Pm1}
\end{equation}
equivalently,
\begin{equation}\label{eq:winding-from-Pm}
q(\boldsymbol{u}_\ast)=\frac{1}{2\pi}\int_0^{2\pi} \frac{d}{d\theta}\,\arg P_m(\cos\theta,\sin\theta)\, d \theta.
\end{equation}
This is the general local relation between null order and winding number. The null order $m$ tells us the degree of the first nonzero term in the Taylor expansion. The winding number $q$ is the winding of that leading term around the origin in the complex plane. Although $m$ and $q$ are are not the same, they are related.

There is a useful exact way to see what values of $q$ are possible. Introduce the complex local coordinates
\begin{equation}
z:=\xi+i\eta, \qquad \bar z:=\xi-i\eta .
\end{equation}
Any homogeneous degree-$m$ polynomial in $(\xi,\eta)$ can be written as
\begin{equation}
P_m(z,\bar z)=\sum_{k=0}^{m}c_k z^k\bar z^{\,m-k},
\label{eq:Pm-z-zbar}
\end{equation}
with complex coefficients $c_k$. On the unit circle,
$z=e^{i\theta}$, thus
\begin{align}
    P_m(e^{i\theta},e^{-i\theta})=\sum_{k=0}^m c_k e^{i(2k-m)\theta}=e^{-im\theta}\sum_{k=0}^m c_k e^{2ik\theta}.
\end{align}
Then 
\begin{equation}
P_m(e^{i\theta},e^{-i\theta}) = e^{-im\theta}p\,(e^{2i\theta}).
\label{eq:pm-factor-circle}
\end{equation}
The first factor contributes to a phase change is $\int_0^{2\pi}(-im)\,d \theta=-2\pi i m$, while the second one is polynomial evaluated at $e^{2i\theta}$. As $\theta$ runs from $0$ to $2 \pi$, $e^{2i\theta}$ goes around the unit circle twice. Therefore, if $p(e^{2i\theta})$ has no zeros on $|e^{2i\theta}|=1$, its phase changes as $4 \pi i N_{\rm in}$, where $N_{\rm in}$ is the number of zeros of $p(e^{2i\theta})$ inside the unit disk\footnote{This is given by the argument principle theorem which roughly states when a complex function has no zeros or poles on a closed contour, the phase change of the function along that contour counts the number of zeros inside the contour minus the number of poles inside, with multiplicity. In our case, for a polynomial there are no poles, so the phase winding counts the zeros.}. Both terms together give 
\begin{equation}
    \int_0^{2 \pi}\frac{\partial_\theta P_m}{P_m}\, d \theta=2 \pi  i\left(- m + 2N_{\rm in}\right).
    \label{eq:q-2n-minu}
\end{equation}
Therefore 
\begin{equation}
    q=- m + 2N_{\rm in}.
\label{eq:q-2n-minu-final}
\end{equation}
Since $0\le N_{\rm in}\le m$ the possible charges at order $m$ are
\begin{equation}\label{eq:possible-winding}
   |q|\leq m, \qquad  q\in \{-m, -m+2, \dots, m-2, m\},
\end{equation}
The most common example is a simple null where visibility is zero but its gradients remains nonzero. Therefore the first nonzero term in the Taylor expansion \autoref{eq:V-homo-expansion} is $P_1(\xi,\eta)$ which defines a real linear map from local $(\xi,\eta)-$plane to the complex plane by $(\xi,\eta) \mapsto (\Re P_1, \Im P_1)$. Therefore if the determinat of Jacobian of this map is nonzero, $\det J \neq 0$, the null is isolated and simple. This means a small circle around the null is mapped to an ellipse around the origin in the complex plane, and thus the phase winds exactly once. The sign of the winding is the orientation of the map whether the phase of $V$ increases clockwise or counterclockwise as one goes counterclockwise around the null in the $(u,v)-$ plane
\begin{equation}
    q(\boldsymbol{u_\ast})=\mathrm{sign} \det J = \pm 1 
\end{equation}
Therefore, every nondegenerate linear null has winding number $+1$ or $-1$. Another restricted example is
\begin{equation}
P_m(\boldsymbol \xi)=(a\cdot\boldsymbol \xi)^m
\end{equation}
If $a\cdot\boldsymbol \xi$ is a nondegenerate complex linear coordinate, then it winds once around the origin as the loop circles the null. Raising it to the $m$-th
power makes the phase wind $m$ times, so
\begin{equation}
q=\pm m.
\end{equation}
However, a general degree-$m$ null need not have winding $\pm m$. For example consider the quadratic case
\begin{equation}
   P_2(\xi,\eta)=(\xi +i \eta)^2. 
\end{equation}
On a small circle $\xi=\epsilon \cos \theta$ and $\eta=\epsilon \sin \theta$ we have $\xi+i\eta=\epsilon e^{i \theta}$. Therefore 
\begin{equation}
    P_2= \epsilon^2 e^{2 i \theta}.
\end{equation}
As $\theta$ runs from $0$ to $2 \pi$, the phase of $P_2$ runs from $0$ to $4\pi$. Therefore 
\begin{equation}
q=2.
\end{equation}
Now consider
\begin{equation}
   P_2(\xi,\eta)=\xi^2 + \eta^2. 
\end{equation}
on the same circle we have
\begin{equation}
    P_2= \epsilon^2.
\end{equation}
This is real and positive for every $\theta$, therefore its phase is constant and $q=0$.
Both examples have null of order $m=2$, but with two different winding numbers. Thus, the null order can change, but the total winding inside a zero free boundary cannot change and is the robust topological quantity. 

Now consider a small region $D$ in the $(u,v)$-plane whose boundary does not pass through a visibility zero. Then the total winding of all nulls inside $D$ is given by
\begin{equation}
Q_D =\sum_{\boldsymbol u_\ast\in D}q(\boldsymbol u_\ast)=\frac{1}{2\pi}\oint_{\partial D}d\,\arg V .
\label{eq:total-winding-domain}
\end{equation}
The right hand side depends only on the phase of $V$ on the boundary of this region $\partial D$. Therefore, under a small continuous perturbation of the image, that no zero crosses $\partial D$, the integer $Q_D$ cannot change. 
This gives a local conservation law that must be obeyed under such perturbations,
\begin{equation}
    Q_D = \sum_{i\in D} q_i = \text{const}.
\end{equation}
Thus, for example, if initially we have a null of charge +2, it can evolve under the image perturbation into two nulls of charge +1 (both enclosed within $D$) but never into two nulls of charge +1 and -1.  

This conservation implies that nulls must appear in locally conserved sets.  We could imagine a sequence of images beginning from a Gaussian and smoothly evolving into the desired intensity map, $I(\boldsymbol{x})$.  Because for any small region $D$, $Q_D=0$ for the original Gaussian, at some point, prior to any nulls crossing $\partial D$, each null generated within $D$ must have a counter-balancing partner.  This need not appear in pairs, e.g., three nulls could appear with $q=+2$, $-1$, and $-1$, though commonly does.

Moreover, the above conservation law is the precise reason in which nulls are created and destroyed in pairs. A single $(+1)$ null cannot suddenly disappear in the middle of a zero free region, because then that would change the boundary winding number by one. Therefore, the generic local creation or annihilation event involves two nulls with opposite charges. 

In addition to this local balancing property of nulls exists a distinct global balance that arises from the hermitian nature of the visibility map.  That is, if $V(\boldsymbol{u_0})=0$, then $V(-\boldsymbol{u_0})=V*(\boldsymbol{u_0})=0$ as well. Moreover, the two charges are opposite $q(-\boldsymbol{u_0})=-q(\boldsymbol{u_0})$, since reflection through the origin $\boldsymbol{u} \mapsto -\boldsymbol{u}$ preserves orientation in the $(u,v)-$plane but the complex conjugate reverses the phase winding; therefore, winds in the opposite direction. 




If null is not an isolated point, for example in symmetric sources such as a perfectly circular ring or disk, they can have null curves instead of null points. Of course along this curve the local charge of an individual point can not be defined in the same way, because one cannot go around this point on the curve while keeping $V\neq 0$ on the loop. However, once the symmetry is broken, the null curve is no longer generic. It disappears or splits into isolated zeros following the conservation rule above. 

Pseudo nulls are out of this discussion, since this is a zero of a chosen component, for example $\Re V=0$ but $|V|\neq 0$; therefore, as mentioned earlier, the complex visibility does not vanish, the phase is well defined and there is no local charge. Thus they are not required to occur in topological pairs. 


The phenomenology of the nulls (including the identification of winding numbers, their topology, and associated conservation laws) immediately raises questions about their potential value.  It remains unresolved precisely how much information we know about an underlying image should we possess the locations and winding numbers of all of the visibility nulls, though it does echo similar aspects from complex analysis (where these would correspond to poles and residues).  Nevertheless, we leave these questions for future work and now focus on their practical utilities in the subsequent sections.



\section{Closure phase response to an arbitrary image}
\label{sec:functionalkernel}

This section develops the central response theory. We first derive a universal visibility space response valid on any zero free branch. We then specialize the same response to two physically distinct situations: an arbitrary image perturbation at a fixed triangle, and a fixed image sampled by a moving triangle. These give the image kernel and moment hierarchy in Subsections \ref{sec:master-functional-kernel-I}-\ref{sec:moment-hierarchy} and the Fourier plane motion hierarchy in Subsection \ref{sec:visib-delta-u}.

\subsection{Closure phase as a log-visibility observable}
\label{sec:log-visibility}

Closure phase is most naturally understood in visibility space. In this subsection the image is fixed, unless stated otherwise, and we ask how the closure phase changes when the sampled visibilities change. For the image $I(\boldsymbol x)$, we use \autoref{eq:visibility-fixed-image}, and consider a closed triangle, 
and present the three sampled visibilities by
\begin{equation}\label{eq:ref-visibility}
V_a := V_I(\boldsymbol u_a), \qquad a=1,2,3 .
\end{equation}
Assume first that none of these visibilities is zero. Then each $V_a$ has a nonzero amplitude and a well-defined phase. Along any zero free part of the observation we can write
\begin{equation}
V_a=|V_a|\,e^{i\theta_a},
\end{equation}
where the phases $\theta_a$ are followed continuously. The continuously followed closure phase is then given by
\begin{equation}
\widetilde\phi_\triangle =\theta_1+\theta_2+\theta_3 .
\label{eq:cp-phase-sum}
\end{equation}
The tilde is used to emphasize that this is the unwrapped closure phase, not a value forced into a fixed interval such as $(-\pi,\pi]$ as the same notation as in \autoref{sec:topolgy-thm}.

It is useful to express the same statement using the logarithm. On a zero free interval we may choose the logarithm consistently with the continuous phases above,
\begin{equation}
\Log V_a=\ln |V_a|+i\theta_a .
\end{equation}
Therefore
\begin{equation}
\widetilde\phi_\triangle=\Im\sum_{a=1}^{3}\Log V_a=\Im\Log(V_1V_2V_3).
\label{eq:cp-logV}
\end{equation}
This is the basic local form of closure phase away from nulls. It says that closure phase is controlled by logarithmic changes of the three complex visibilities.

We now want to explore: how much does the closure phase change when the three complex visibilities are changed by finite, but controlled, amounts? 

We therefore start by writing

\begin{equation}
V_a' = V_a+\Delta V_a ,
\end{equation}
with no visibility crossing zero during the change. The new unwrapped closure phase is
\begin{equation}
\widetilde\phi_\triangle'=\Im\sum_{a=1}^{3}\Log V_a' .
\end{equation}
Since $V_a+\Delta V_a=V_a\left(1+\Delta V_a/V_a\right)$,
we obtain
\begin{equation}\label{eq:finite-dlogV}
\Delta\widetilde\phi_\triangle=\widetilde\phi_\triangle'-\widetilde\phi_\triangle=\Im\sum_{a=1}^{3}\Log\left(1+\frac{\Delta V_a}{V_a}\right).
\end{equation}
Note that this is not a new assumption; it is just the closure phase written in terms of the fractional changes of the three visibilities. For small changes, we can write $\Log(1+z)=z+\mathcal O(z^2)$; therefore
\begin{equation}\label{eq:cp-dV-over-V}
\delta\widetilde\phi_\triangle=\Im\sum_{a=1}^{3}\frac{\delta V_a}{V_a}.
\end{equation}
This is the central local response formula in visibility space. The natural variable is not $\delta V_a$ itself, but the fractional complex change $\delta V_a/V_a$. This also makes clear why closure phase becomes sensitive near a visibility null: if one $|V_a|$ is small, a small absolute change in that visibility can produce a large phase response. This is precisely the fork regime that is discussed in \autoref{sec:closure phase-forks}. On such a branch, finite changes are controlled by \autoref{eq:finite-dlogV}, and the first order response is the fractional visibility identity
\autoref{eq:cp-dV-over-V}. We use this visibility space viewpoint to state the exact invariances of closure
phase in the next subsection.

\subsection{Exact visibility space invariances}
\label{sec:visibility-space-invariances}

In this subsection, we aim to record the finite transformation to which closure phases are not sensitive. While these are typically established using the standard proof, we instead present a derivation in visibility space and show infinitesimal versions of these transformations give zero closure phase response. 
For the three baselines of a triangle we write the sample visibilities as $V_a$ for $a=1,2,3$. On any part of the path where all $V_a$ remain nonzero, the unwrapped closure phase is given by \autoref{eq:cp-logV}. 

Now we assume an operation changes the image from $I$ to $I'$ also the three visibilities do not cross zero during this change. Then
\begin{equation}
\widetilde\phi_\triangle[I']-  \widetilde\phi_\triangle[I]=\Im \sum_{a=1}^{3}\Log\left(\frac{V_a[I']}{V_a[I]}\right),
\end{equation}
this is the only identity that we need in this subsection.
The important special case is when an operation acts multiplicatively on the three visibilities as
\begin{equation}
    V_a[I']=m_a V_a[I],
\end{equation}
Then 
\begin{equation}
\widetilde\phi_\triangle[I']-  \widetilde\phi_\triangle[I] = \Im\Log\,(m_1\,m_2\,m_3).
\label{eq:multi-cp-invar}
\end{equation}
Then the closure phase is unchanged whenever the product $(m_1\,m_2\,m_3)$ is real and positive.

\subsubsection{Positive global rescaling}

If $g>0$ and rescale the image by
\begin{equation}
I_g(\boldsymbol x):=g\,I(\boldsymbol x).
\end{equation}
Then every visibility is multiply with the same positive real number
\begin{equation}
  V_a[I_g]=g\,V_a[I],  
\end{equation}
and since $(m_1\,m_2\,m_3)=g^3>0$ we obtain
\begin{equation}
 \widetilde\phi_\triangle[I_g]-  \widetilde\phi_\triangle[I] = \Im\Log\,(g^3) =0,  
\end{equation}
equivalently
\begin{equation}\label{eq:gai-invar-logV}
 \phi_\triangle[I_g]=\phi_\triangle[I]   \qquad (\mathrm{mod} \ 2\pi).
\end{equation}
Therefore, closure phase is insensitive to the flux normalization of the image. 

\subsubsection{Image Translation}
If the image be shifted by an angular displacement $\boldsymbol\delta$ as
\begin{equation}
I_{\boldsymbol\delta}(\boldsymbol x)=I(\boldsymbol x-\boldsymbol\delta).
\end{equation}
With the Fourier convention used here, the shift theorem gives
\begin{equation}
V_a[I_{\boldsymbol\delta}]=e^{2\pi i\,\boldsymbol u_a\cdot\boldsymbol\delta}V_a[I].
\label{eq:visibility-translation-factor}
\end{equation}
The three multiplicative factors are therefore pure phases and their product is
\begin{equation}
\prod_{a=1}^{3}e^{2\pi i\,\boldsymbol u_a\cdot\boldsymbol\delta}
=\exp\left[2\pi i\left(\sum_{a=1}^{3}\boldsymbol u_a\right)\cdot\boldsymbol\delta\right].\\[4pt]
\end{equation}
Since $\sum_{a=1}^{3}\boldsymbol u_a=0$, this product is exactly one. Therefore
\begin{equation}
\phi_\triangle[I_{\boldsymbol\delta}]
=\phi_\triangle[I]
\qquad(\mathrm{mod}\ 2\pi).
\label{eq:translation-invariance-logV}
\end{equation}
A translation rotates the three individual visibility phases, but the rotations cancel around a closed triangle. Therefore, closure phase is independent of the actual position of the image.

\subsubsection{Real and positive convolution blur}
Thus closure phase is invariant not only under circular Gaussian smoothing, but also under centered elliptical Gaussian smoothing. The blur may suppress different baseline directions by different amounts, but it does not rotate any complex visibility. Since closure phase depends only on the phases of the three visibility factors, not on their positive real amplitudes, it remains unchanged. 

Now as an example consider the image be convolved with a circular kernel,
\begin{equation}
I_K=K*I, \qquad K(\boldsymbol x)=K(\|\boldsymbol x\|).
\end{equation}
Then the visibility is multiplied by the transfer function of the blur. For a real circular kernel, $\widehat K$ is real and depends only on $\rho_a=\|\boldsymbol u_a\|$, so
\begin{equation}
V_a[I_K]=\widehat K(\rho_a)V_a[I].
\label{eq:blur-visibility-factor-logV}
\end{equation}
If the three transfer factors sampled by the triangle are positive,
\begin{equation}
\widehat K(\rho_1)>0, \qquad \widehat K(\rho_2)>0,
\qquad \widehat K(\rho_3)>0,
\end{equation}
then their product is positive. Therefore
\begin{equation}
\widetilde\phi_\triangle[I_K]- \widetilde\phi_\triangle[I]
=\Im\Log\left[\widehat K(\rho_1)\widehat K(\rho_2)\widehat K(\rho_3)\right]=0,
\end{equation}
equivalently
\begin{equation}
\phi_\triangle[I_K]=\phi_\triangle[I]
\qquad(\mathrm{mod}\ 2\pi).
\label{eq:blur-invariance-logV}
\end{equation}
A real and positive circular blur changes the visibility amplitudes, but it does not rotate their product. The closure phase is therefore unchanged. A Gaussian blur is the simplest example of this case, since its transfer function is positive. In general, for any convolution kernel $K$ that $\widehat K(\boldsymbol u_{ij})$, $\widehat K(\boldsymbol u_{jk})$, and $\widehat K(\boldsymbol u_{ki})$ are positive and real, one has
\begin{equation}
\phi_\triangle[K*I]=\phi_\triangle[I]\qquad(\mathrm{mod}\ 2\pi).
\end{equation}
If the transfer factors are real but their product is negative, the closure phase shifts by $\pi$.If any transfer factor vanishes, then the blurred bispectrum vanishes and the closure phase is undefined.

\medskip

We now use the same visibility space response formula as in \autoref{sec:log-visibility} in a different way. More preciesly, we have so far treated $\delta V_a$ as an abstract change in the three complex visibilities. The responce formula \autoref{eq:cp-dV-over-V} does not  specify why the visibilities changed. This is useful, because there are two physically different ways that such a change can arise. The baselines may be fixed while the image is changed, or the image may be fixed while the projected baselines move through the Fourier plane. We explore both cases in the \autoref{sec:master-functional-kernel-I} and \autoref{sec:visib-delta-u}.

\subsection{Image perturbations and the master kernel}
\label{sec:master-functional-kernel-I}

The first case is the one needed for the source problem. In comparing images, the sampled triangle may be held fixed while the brightness distribution changes. This is the relevant question when one asks which part of an image controls a closure phase, how a small substructure changes the data, or how two nearby source models differ at the same baselines. In that case the visibility change is not caused by the Fourier component of the image perturbation, that we now discuss this case precisely.


Let take $I_0(\boldsymbol x)$ as a reference image, and perturb it by
\begin{equation}
I_\epsilon(\boldsymbol x)=I_0(\boldsymbol x)+\epsilon\,\delta I(\boldsymbol x),\qquad|\epsilon|\ll1,
\end{equation}
where $\delta I(\boldsymbol x)$ is an arbitrary real brightness perturbation. The three baselines of the triangle are denoted by $\boldsymbol u_1$ ,$\boldsymbol u_2$, and $\boldsymbol u_3$ where $\boldsymbol u_1+\boldsymbol u_2+\boldsymbol u_3=0$. All quantities in this subsection are evaluated at these fixed baselines. We write the reference visibilities as \autoref{eq:ref-visibility} and assume $V_1^0V_2^0V_3^0\neq0$.
So that, the reference triangle is near a fork/null, and the unwrapped closure phase can be linearized.

The perturbation $\delta I$ changes the visibility on the $a$-th baseline by the Fourier component of $\delta I$ sampled at that baseline:
\begin{equation}\label{eq:dV-from-dI}
\delta V_a=\left.
\frac{d}{d\epsilon}V_{I_\epsilon}(\boldsymbol u_a)
\,\right|_{\epsilon=0}=\int_{\mathbb R^2}
\delta I(\boldsymbol x)\,
e^{2\pi i\,\boldsymbol u_a\cdot\boldsymbol x}
\,d^2\boldsymbol x.
\end{equation}
The visibility space response law derived \autoref{eq:cp-dV-over-V} gives
\begin{equation}
\delta\widetilde\phi_\triangle=\Im\sum_{a=1}^{3}
\frac{\delta V_a}{V_a^0}.
\label{eq:cp-dV-over-V-image}
\end{equation}
Substituting \autoref{eq:dV-from-dI} into this expression gives the closure phase response as a single integral over the image
\begin{equation}\label{eq:kernel-response}
\delta\widetilde\phi_\triangle=\int_{\mathbb R^2}
\mathscr K_{\triangle,I_0}(\boldsymbol x)\,\delta I(\boldsymbol x)\, d^2\boldsymbol x.
\end{equation}
The weight function
\begin{equation}
\mathscr K_{\triangle,I_0}(\boldsymbol x)=
\Im\left[\frac{e^{2\pi i\,\boldsymbol u_1\cdot\boldsymbol x}}{V_1^0}+\frac{e^{2\pi i\,\boldsymbol u_2\cdot\boldsymbol x}}{V_2^0}+\frac{e^{2\pi i\,\boldsymbol u_3\cdot\boldsymbol x}}{V_3^0}\right]
\label{eq:master-kernel}
\end{equation}
is the master functional kernel. It gives the first order change in closure phase caused by adding a small amount of brightness at position $\boldsymbol x$. In this sense $\mathscr K_{\triangle,I_0}$ is a leverage map for the closure phase of the chosen triangle.

The formula is often most transparent after writing each reference visibility in amplitude phase form,
\begin{equation}
V_a^0=|V_a^0|\,e^{i\psi_a}.
\end{equation}
Then the contribution from the $a$-th baseline is
\begin{equation}
\Im\left[ \frac{e^{2\pi i\,\boldsymbol u_a\cdot\boldsymbol x}}{V_a^0}\right]=\frac{1}{|V_a^0|}\sin\left(2\pi\,\boldsymbol u_a\cdot\boldsymbol x-\psi_a\right),
\label{eq:kernel-baseline-sine}
\end{equation}
and therefore
\begin{equation}
\mathscr K_{\triangle,I_0}(\boldsymbol x)=\sum_{a=1}^{3}
\frac{1}{|V_a^0|}\sin\left(2\pi\,\boldsymbol u_a\cdot\boldsymbol x-\psi_a\right).
\label{eq:master-kernel-sine}
\end{equation}
Therefore, the kernel is a sum of three baseline response terms. Each term oscillates across the image with the spatial frequency of one baseline and is weighted by the inverse visibility amplitude on that baseline.

This inverse amplitude weighting is the practical content of the formula. If one of the three reference visibilities is small, then a small image perturbation can produce a large closure phase response. In particular,
\begin{equation}
|\mathscr K_{\triangle,I_0}(\boldsymbol x)|
\le \sum_{a=1}^{3} \frac{1}{|V_a^0|}.
\label{eq:kernel-bound-pointwise}
\end{equation}
Hence, for any perturbation with finite total absolute flux,
\begin{equation}
|\delta\widetilde\phi_\triangle|\le
\|\delta I\|_1 \sum_{a=1}^{3} \frac{1}{|V_a^0|}, \qquad 
\label{eq:kernel-response-bound}
\end{equation}
where $\|\delta I\|_1 :=\int_{\mathbb R^2}|\delta I(\boldsymbol x)|\,d^2\boldsymbol x.$

If $\rho_\triangle:=\min_{a=1,2,3}|V_a^0|$, then \autoref{eq:kernel-response-bound} becomes
\begin{equation}\label{eq:rho-amplification}
|\delta\widetilde\phi_\triangle|\le\frac{3\,\|\delta I\|_1}{\rho_\triangle}.
\end{equation}
This is the image perturbation version of the null amplification discussed earlier. As a baseline visibility approaches a true null, the closure phase becomes increasingly sensitive to small changes in the source.

At the null itself the kernel is not defined. If one of the three $V_a^0$ vanishes, then the reference bispectrum vanishes, the closure phase is undefined, and the linear response of the phase cannot be formed at that point. Thus \autoref{eq:master-kernel} describes the
regular response away from forks, while its divergence signals the approach to the fork regime.

The kernel also reflects the exact invariances discussed in the previous subsection. For an infinitesimal positive rescaling of the image, $\delta I=\eta I_0$, one has $\delta V_a=\eta V_a^0$, and therefore
\begin{equation}
\delta\widetilde\phi_\triangle=\Im\sum_{a=1}^{3}\eta=0.
\end{equation}
For an infinitesimal translation of the image by $\boldsymbol\delta$, the induced visibility change is
\begin{equation}
\delta V_a=2\pi i\,(\boldsymbol u_a\cdot\boldsymbol\delta)\,V_a^0 .
\end{equation}
Thus
\begin{equation}
\delta\widetilde\phi_\triangle=\Im\left[2\pi i\left(\sum_{a=1}^{3}\boldsymbol u_a\right)\cdot\boldsymbol\delta\right]=0,
\end{equation}
because the baselines close. These checks are useful: the master kernel responds to changes in image structure, not to an overall flux scale or to an absolute shift of the image on the sky.


\subsection{Moment hierarchy for image perturbations}
\label{sec:moment-hierarchy}


The master kernel gives the exact first order response to an arbitrary brightness perturbation. It is also useful to expand this response in ordinary moments of $\delta I$. This gives a hierarchy: the closure phase response can be decomposed into the response to the total added flux, the dipole moment of the perturbation, the quadrupole moment, and so on.

We keep the same fixed triangle and reference image as in the previous subsection. Starting from the master kernel \autoref{eq:kernel-response}, we expand each exponential
\begin{equation}
e^{2\pi i\,\boldsymbol u_a\cdot\boldsymbol x}=\sum_{N=0}^{\infty}\frac{(2\pi i)^N}{N!}
(\boldsymbol u_a\cdot\boldsymbol x)^N .
\end{equation}
The $N$-th ordinary moment of the perturbation is
\begin{equation}\label{eq:perturbation-moments}
M^{(N)}[\delta I]:=
\int\delta I(\boldsymbol x)\,\boldsymbol x^{\otimes N}\, d^2\boldsymbol x,
\end{equation}
with $M^{(0)}[\delta I]=\int \delta I(\boldsymbol x)d^2\boldsymbol x$. 

The powers of $\boldsymbol x$ can be collected into of the perturbation. To do this compactly, we write $\boldsymbol x^{\otimes N}$ for the $N$-fold tensor product of $\boldsymbol x$, and use $:$ for contraction over all tensor indices. With this
notation,
\begin{equation}
(\boldsymbol u_a\cdot\boldsymbol x)^N=\boldsymbol u_a^{\otimes N}:\boldsymbol x^{\otimes N}.
\end{equation}
The first cases are just the familiar ones: for $N=1$ this is the ordinary dot product, and for $N=2$ it is the quadratic form 
\begin{equation}
\boldsymbol u_a^{\otimes 2}:\boldsymbol x^{\otimes 2}=
\sum_{\alpha,\beta}u_{a,\,\alpha}\,u_{a,\,\beta}\,x_\alpha x_\beta =(\boldsymbol u_a\cdot\boldsymbol x)^2.
\end{equation}
Therefore, the response becomes
\begin{equation}
\delta\widetilde\phi_\triangle
=\sum_{N=0}^{\infty}(2\pi)^N\mathcal K_\triangle^{(N)}:
M^{(N)}[\delta I],
\label{eq:all-moments}
\end{equation}
and the coefficient of the $N$-th perturbation moment is\footnote{Assuming that $\delta I$ has finite total absolute flux and is supported within a bounded field of view, the moment series converges absolutely at every finite baseline; an explicit truncation bound is given in \autoref{app:conve}.}
\begin{equation}
\mathcal K_\triangle^{(N)}=\frac{1}{N!}
\Im\left[i^N\sum_{a=1}^{3}
\frac{\boldsymbol u_a^{\otimes N}}{V_a^0}\right],
\qquad\boldsymbol u_a^{\otimes0}:=1 .
\label{eq:moment-kernel-coefficients}
\end{equation}
Here the imaginary part acts component by component on the complex tensor inside the brackets. The baselines $\boldsymbol u_a$ are the actual sampled baseline coordinates of the triangle.
The first few terms are
\begin{align}
\delta\widetilde\phi_\triangle=&\,
\mathcal K_\triangle^{(0)} M^{(0)}[\delta I]+
2\pi\,\mathcal K_\triangle^{(1)}:M^{(1)}[\delta I]\nonumber\\
&+(2\pi)^2\,\mathcal K_\triangle^{(2)}:M^{(2)}[\delta I]+\cdots.
\label{eq:moment-hierarchy-first-terms}
\end{align}
In fact, each closure phase acts as a linear functional on the moment tower of the perturbation. The $N=0$ term is the response to the rank zero, or monopole, part of the perturbation; the $N=1$ term is the response to its rank one moment; the $N=2$ term is the response to its rank two Cartesian moment, and so on.

This hierarchy should be read as a rank $N$ Cartesian moment expansion. In two image dimensions, a rank $N$ Cartesian moment need not be a pure angular multipole. For example, a rank two moment contains both a trace part and a quadrupolar part. If one wants to have a pure dipole, quadrupole, octupole, etc., should further decompose the Cartesian tensors into their irreducible angular pieces. However, for the present result the rank hierarchy is the natural form because it follows directly from the Taylor expansion of the visibility kernel.

The useful point is that all dependence on the sampled triangle and on the reference image is contained in the known tensors $\mathcal K_{\triangle,I_0}^{(N)}$, while all dependence on the perturbation is contained in the moments $M^{(N)}[\delta I]$. Thus, this formula shows the closure phase separated into a geometric/sensitivity part and a source structure part. Truncating the series gives a controlled low order description of which spatial moments of an image perturbation are visible to a given triangle.
If the perturbation conserves total flux, then
\begin{equation}
M^{(0)}[\delta I]=0,
\end{equation}
and the monopole term drops out. If, in addition, the perturbation has no rank one moment about the chosen image origin, then the leading possible response may begin at rank two order. This is often the useful way to interpret the expansion: once the lower moments vanish, the first nonzero contraction in the hierarchy identifies the leading moment order of the image change seen by that closure phase.

The full series cannot replace the master kernel in all regimes. It is most useful when perturbation is supported in a bounded region and the first few moments give a proper approximation.

 \subsection{Closure phase response to motion in Fourier plane}\label{sec:visib-delta-u}

The moment hierarchy in \autoref{sec:master-functional-kernel-I} described how closure phase responds when the image changes at fixed baselines. In this subsection, We consider the complementary situation. The image $I_0(\boldsymbol x)$ is held fixed, while Earth rotation moves the observing triangle through the Fourier plane. More generally, the same description applies whenever the projected baselines or the observing wavelength change.

This viewpoint has a useful practical advantage. At the reference triangle, the three complex visibilities \autoref{eq:ref-visibility} are known from the data or from a reference image. We do not need to reconstruct these reference values from a moment expansion. Instead, we keep $V_a^0$ exact and use the image moments only to describe how the visibilities change when the triangle moves away from the reference point.


 \smallskip
 
A small displacement $\boldsymbol u_a\to \boldsymbol u_a+\delta\boldsymbol u_a$, with the triangle kept
closed, changes the visibility sampled on the $a$-th baseline from $V_a^0=V_{I_0}(\boldsymbol u_a)$ to
$V_{I_0}(\boldsymbol u_a+\delta\boldsymbol u_a)$. We use the same Cartesian moments \autoref{eq:perturbation-moments}, evaluated for the fixed image $I_0$. Therefore, expanding the Fourier kernel gives the visibility in terms of these moments gives
\begin{equation}
V_{I_0}(\boldsymbol u)=\sum_{N=0}^{\infty}
\frac{(2\pi i)^N}{N!}\, \boldsymbol u^{\otimes N}\,:\,M^{(N)}[I_0].
\label{eq:visib-image-moment}
\end{equation}
This is the same expansion that produced the perturbation hierarchy, but it is now applied to the reference image itself. For an image of bounded extent the series converges at every finite baseline. A truncation at rank $N_{\max}$ gives an explicit polynomial of degree $N_{\max}$ in the Fourier plane coordinates.

Evaluating \autoref{eq:visib-image-moment} at the displaced baseline and subtracting the reference value gives
\begin{equation}\label{eq:delta-u-visibi}
\Delta V_a^{\rm geom} = \sum_{N=1}^{\infty}
\frac{(2\pi i)^N}{N!}\left[(\boldsymbol u_a+\delta\boldsymbol u_a)^{\otimes N}-
\boldsymbol u_a^{\otimes N} \right]\,:\,M^{(N)}[I_0].
\end{equation}
The rank zero term cancels because moving the baseline does not change the total flux contribution. At each rank, the quantity in square brackets is a finite polynomial in the reference baseline and its displacement. For example,
\begin{equation}
(\boldsymbol u_a+\delta\boldsymbol u_a)^{\otimes2}
-\boldsymbol u_a^{\otimes2}=\boldsymbol u_a\otimes\delta\boldsymbol u_a+ \delta\boldsymbol u_a\otimes\boldsymbol u_a+\delta\boldsymbol u_a^{\otimes2}.
\end{equation}
Therefore, the same image moments that organize the response to $\delta I$ also organize the variation of the visibility. As long as none of the three visibilities crosses a null during the displacement, the finite change in the unwrapped closure phase is given by \autoref{eq:finite-dlogV}. Together, with \autoref{eq:delta-u-visibi} give the finite, zero free response of closure phase to motion of the triangle. The visibility change is polynomial at every finite moment
order; the logarithm converts that visibility change into the corresponding phase change.
For a sufficiently small displacement, only terms that are linear in $\delta\boldsymbol u_a$ are needed. At rank $N$, differentiating $\boldsymbol u_a^{\otimes N}$ replaces one of its $N$ baseline factors by $\delta\boldsymbol u_a$. Since $M^{(N)}[I_0]$ is symmetric, all $N$ choices give the
same contraction. Therefore
\begin{equation}\label{eq:linear-delta-u-visib}
\delta V_a^{\rm geom}=\sum_{N=1}^{\infty}\frac{(2\pi i)^N}{(N-1)!}\left(\boldsymbol u_a^{\otimes(N-1)}
\otimes \delta\boldsymbol u_a\right)\,:\,M^{(N)}[I_0].
\end{equation}
Substituting this result into $\delta\widetilde\phi_\triangle=\Im\sum_a\delta V_a/V_a^0$ gives the geometric moment hierarchy
\begin{align}\label{eq:geo-hierarchy}
\delta\widetilde\phi_\triangle^{\rm geom}=\Im\sum_{a=1}^{3} \sum_{N=1}^\infty \frac{(2\pi i)^N}{(N-1)!V_a^0}\, \left(\boldsymbol{u_a}^{\otimes (N-1)}\otimes \delta \boldsymbol{u_a} \right) \, : \, M^{(N)}[I_0]
\end{align}
Equivalently
\begin{equation}
\delta\widetilde\phi_\triangle^{\rm geom} =\sum_{N=1}^{\infty}
(2\pi)^N \mathcal G_{\triangle,I_0}^{(N)}: M^{(N)}[I_0],
\label{eq:geo-hierarchy-short}
\end{equation}
where the response tensors
\begin{equation}
\mathcal G_{\triangle,I_0}^{(N)} \,:=\, \frac{1}{(N-1)!} \Im\left[i^N \sum_{a=1}^{3} \frac{ \boldsymbol u_a^{\otimes(N-1)}
\otimes \delta\boldsymbol u_a}{V_a^0}\right],
\label{eq:geo-coefficient}
\end{equation}
contain the reference baselines, their
displacements, and the known reference visibilities.

This result has the same structure as the hierarchy for an image perturbation. There, the moments $M^{(N)}[\delta I]$ described the change in the source, while the triangle was
fixed. Here the moments $M^{(N)}[I_0]$ describe the fixed source, while the factors $\boldsymbol u_a^{\otimes(N-1)}\otimes\delta\boldsymbol u_a$ describe the motion of the
triangle. At rank $N$, the coefficient in the earlier hierarchy contained $\boldsymbol u_a^{\otimes N}/N!$; baseline motion replaces one of those factors by $\delta\boldsymbol u_a$, producing the coefficient
$\boldsymbol u_a^{\otimes(N-1)}\otimes\delta\boldsymbol u_a/(N-1)!$.

The first few terms are
\begin{align}
\delta\widetilde\phi_\triangle^{\rm geom}=\Im\sum_{a=1}^{3}
\Bigg[&\frac{2\pi i}{V_a^0}\,\delta\boldsymbol u_a\cdot M^{(1)}[I_0]\\
&+\frac{(2\pi i)^2}{V_a^0}\,\left(\boldsymbol u_a\otimes\delta\boldsymbol u_a\right):M^{(2)}[I_0]\nonumber\\
&+\frac{(2\pi i)^3}{2V_a^0}\,\left(\boldsymbol u_a^{\otimes2}\otimes\delta\boldsymbol u_a\right):M^{(3)}[I_0]+\cdots \Bigg]\nonumber.
\end{align}
These are again Cartesian rank moments. No decomposition into pure angular multipoles has been made. At any finite rank, \autoref{eq:geo-hierarchy} is a polynomial in the reference baseline coordinates and is linear in the displacement. This is useful when a low rank description of the image is enough: the image moments can be computed once, while the known baseline coordinates and their displacements determine the closure phase variation across the observing track.

For an observing track $\boldsymbol u_a(t)$, setting
$\delta\boldsymbol u_a=\dot{\boldsymbol u}_a\,dt$, \autoref{eq:linear-delta-u-visib} gives


\begin{align}\label{eq:106!}
\frac{d \widetilde\phi_\triangle}{dt}=\Im\sum_{a=1}^{3} \sum_{N=1}^\infty \frac{(2\pi i)^N}{(N-1)!V_a^0}\, \left(\boldsymbol{u_a}^{\otimes (N-1)}\otimes  \boldsymbol{\dot{u}_a}\right) :  M^{(N)}[I_0]
\end{align}
with all quantities evaluated at the time of interest. This separates the source structure, contained in $M^{(N)}[I_0]$, from the motion of the array through the Fourier plane, contained in $\boldsymbol u_a(t)$ and $\dot{\boldsymbol u}_a(t)$\footnote{Note that the linear closure phase response requires the fractional visibility change to be small. Near a simple visibility null, the relevant scale is the distance from the reference baseline to the null. In that regime, this ratio is approximately the displacement divided by this distance, with the precise value depending on the direction of motion. Thus the allowed linear neighborhood shrinks as the baseline approaches a null. The visibility moment expansion itself remains regular at a null. What fails there is the division by $V_a^0$, and therefore the regular expansion of the closure phase. At the null the phase is undefined, and the local behavior is instead described by the closure phase fork.}.

\subsection{Squeezed closure triangles and cross-scale response} \label{sec:squeezed-cosmo}

\begin{figure*}
    \centering
      \includegraphics[width=0.9\linewidth]{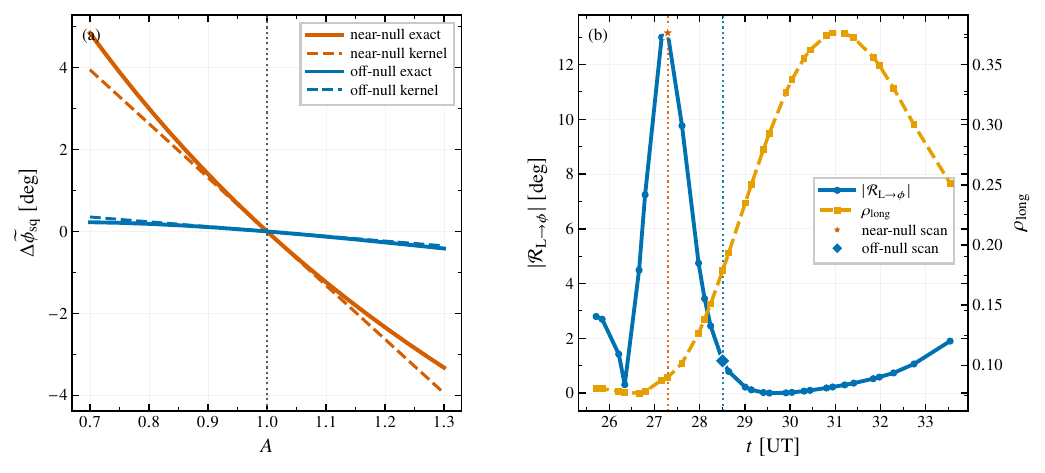}
    \caption{Cross-scale closure phase response evaluated from the best fit Themis image on the actual band 3 in GL-KT-MG track $M87^*$. The triangle is squeezed, with KT-MG forming the short leg and the other two baselines sampling nearby high spatial frequency visibility points. The large scale perturbation $I_{\rm L}$ is obtained by smoothing the inversion odd component of the image with a $20\,\mu{\rm as}$ Gaussian. \textit{Left:} exact finite closure phase change as the amplitude $A$ of $I_{\rm L}$ is varied, compared with the linear master kernel response. The near null configuration is much more sensitive than the off-null configuration, while the departure from the tangent response at larger $|A-1|$ shows when the finite expression becomes necessary. \textit{Right:} magnitude of the cross-scale response coefficient along the Earth rotation track, together with the minimum normalized visibility amplitude of the two long baselines. The response is strongly enhanced while the long baseline correlated flux is suppressed. The marked scans are the two configurations compared in the left panel.}
    \label{fig:cross-scale-cmb-m87}
\end{figure*}
\citet{PajerZaldarriaga2012}
showed that the cross correlation of CMB $\mu$-distortion with temperature anisotropy opens a window onto squeezed primordial non-Gaussianity: the distortion contains the short scale power, while the temperature anisotropy provides the long wavelength mode.
Their observable and closure phase are of course physically different, but their construction suggests a useful operational question for interferometry. Can a change in broad source structure be detected through the closure phase measured by much longer baselines? 

The natural geometry is a squeezed closure triangle. We write
\begin{equation}
\boldsymbol u_1=\boldsymbol q, \quad
\boldsymbol u_2=\boldsymbol k, \quad
\boldsymbol u_3=-\boldsymbol k-\boldsymbol q, \quad \text{ with }\,\,|\boldsymbol q|\ll|\boldsymbol k|.
\label{eq:squeezed-closure-triangle}
\end{equation}
The short baseline $\boldsymbol q$ is sensitive to the broad structure of the source, while the other two baselines sample nearby points in the high spatial frequency visibility. For a real image,
\begin{equation}\label{eq:squeezed-bispectrum}
 B_{\rm sq}(\boldsymbol q,\boldsymbol k)=V(\boldsymbol q)\,
V(\boldsymbol k)\, V^\ast(\boldsymbol k+\boldsymbol q),
\end{equation}
and therefore, on a zero free branch,
\begin{equation}
\widetilde\phi_{\rm sq}=\phi(\boldsymbol q)+\phi(\boldsymbol k)-\phi(\boldsymbol k+\boldsymbol q),
\label{eq:squeezed-cp}
\end{equation}
where $\phi(\boldsymbol u):=\arg V(\boldsymbol u)$. This expression has a simple interpretation. The two long baselines form a finite difference of the small scale visibility phase, while the short baseline gives the
large scale phase needed to make that difference translation invariant. More precisely, the difference of the small scale alone, depends on the arbitrary choice of image origin. The short baseline supplies the missing low spatial frequency phase. Under a translation of the image, its phase changes by
exactly the amount needed to cancel the residual translation phase of the two long baselines. For a sufficiently short baseline this contribution is controlled at leading
order by the image centroid. A squeezed closure triangle therefore measures the local high-frequency phase variation relative to the large scale center of the source, rather than relative to an arbitrary coordinate system. Since $|\boldsymbol q|\ll|\boldsymbol k|$, we expand
$\phi(\boldsymbol k+\boldsymbol q)$ about $\boldsymbol k$, which gives
\begin{equation}\label{eq:squeezed-cp-gradient}
\widetilde\phi_{\rm sq} = \phi(\boldsymbol q)
- \boldsymbol q\cdot \nabla_{\boldsymbol k}\phi(\boldsymbol k) + \mathcal O(q^2).
\end{equation}
If $F$ and $\boldsymbol x_{\rm c}$ are the total flux and brightness centroid, then
\begin{equation}
\phi(\boldsymbol q) = 2\pi\boldsymbol q\cdot\boldsymbol x_{\rm c} + \mathcal O(q^3),
\end{equation}
therefore 
\begin{equation}
\widetilde\phi_{\rm sq} = - \boldsymbol q\cdot \left[ \nabla_{\boldsymbol k}\phi(\boldsymbol k) - 2\pi\boldsymbol x_{\rm c} \right] + \mathcal O(q^2).
\label{eq:squeezed-cp-invariant-gradient}
\end{equation}
The combination in brackets is independent of the arbitrary image origin. A translation adds the same vector $2\pi\boldsymbol c$ to both $\nabla_{\boldsymbol k}\phi$ and $2\pi\boldsymbol x_{\rm c}$, leaving the squeezed closure phase unchanged. \autoref{eq:squeezed-cp-invariant-gradient} shows that a squeezed triangle probes a directional derivative of the high frequency visibility phase. This immediately makes
the construction sensitive to visibility nulls. Near a null of either long baseline visibility,
\begin{equation}
\nabla_{\boldsymbol k}\phi = \Im\nabla_{\boldsymbol k}\Log V,
\end{equation}
can become large, so a short Fourier leg can reveal rapid small scale phase structure through an otherwise modest closure phase signal. We can make the long-short response more explicit by separating the image into a broad component and a fine component. We decompose $I(\boldsymbol x)$ as
\begin{equation}
I(\boldsymbol x) = I_{\rm S}(\boldsymbol x)+I_{\rm L}(\boldsymbol x),
\label{eq:large-small-image-split}
\end{equation}
where $I_{\rm L}$ is the extended structure image component selected by a fixed and explicitly specified smoothing prescription, and $I_{\rm S}=I-I_{\rm L}$ contains the remaining
structure. Note that the separation into large scale and small scale structure is not unique, so one must choose and state a definite rule for making it. The filter and its characteristic scale are held fixed throughout the calculation. We then vary the strength of this component while leaving its shape fixed
\begin{equation}
I_A(\boldsymbol x)=I_{\rm S}(\boldsymbol x)+A\,I_{\rm L}(\boldsymbol x), \qquad
V_A(\boldsymbol u)
=V_{\rm S}(\boldsymbol u)+ A\,V_{\rm L}(\boldsymbol u).
\label{eq:large scale-amplitude-family}
\end{equation}
The reference image corresponds to $A=1$. Any overall positive normalization of $I_A$ is not relevant for closure phase. The response of the squeezed closure phase to the broad image component follows directly from the master response law
\begin{equation}
\mathcal R_{\rm L\rightarrow\phi}
(\boldsymbol q,\boldsymbol k):= \left. \frac{\partial\widetilde\phi_{\rm sq}}
{\partial A} \right|_{A=1} =
\Im\sum_{a=1}^{3} \frac{ V_{\rm L}(\boldsymbol u_a)}{
V(\boldsymbol u_a)}.
\label{eq:large-small-c-res}
\end{equation}
For the squeezed triangle in \autoref{eq:squeezed-closure-triangle}, Hermitian symmetry gives the equivalent form
\begin{equation}
\mathcal R_{\rm L\rightarrow\phi}
= \Im \left[ \frac{V_{\rm L}(\boldsymbol q)}{V(\boldsymbol q)}
+\frac{V_{\rm L}(\boldsymbol k)}{V(\boldsymbol k)}-\frac{V_{\rm L}(\boldsymbol k+\boldsymbol q)}{V(\boldsymbol k+\boldsymbol q)}\right].
\label{eq:squeezed-large scale-res}
\end{equation}
This is a cross-scale observable. The same broad image component is sampled directly by the short baseline and through its effect on the two neighboring high frequency visibilities.

Now the image plane meaning is obtained by applying the master kernel \autoref{eq:kernel-response} as
\begin{equation}
\mathcal R_{\rm L\rightarrow\phi}
=\int_{\mathbb R^2} \mathscr K_{\triangle,I}(\boldsymbol x)\,
I_{\rm L}(\boldsymbol x)\,d^2\boldsymbol x.
\label{eq:large scale-master-kernel}
\end{equation}
Thus, the kernel identifies which regions of the broad source component have the greatest leverage on the small scale closure phase signal. The squeezed geometry selects the
long-short coupling, while the kernel gives its spatial origin in the image.

The response becomes large near a visibility minimum because the reference visibility is small. However, the apparent enhancement near a visibility minimum should not be interpreted as an unbounded physical response. When the visibility change produced by the large scale part of the image becomes comparable to the reference visibility, the linear approximation is no longer valid and must be replaced by the finite expression
\begin{equation}
\Delta\widetilde\phi_{\rm sq}
= \Im\sum_{a=1}^{3}\Log \left[ 1+\Delta A\, \frac{
V_{\rm L}(\boldsymbol u_a) }{V(\boldsymbol u_a) } \right].
\label{eq:large scale-finite-res}
\end{equation}
This remains valid while the path from $A=1$ to $A=1+\Delta A$ is zero free. If one of the total visibilities becomes zero during this change, its phase is undefined at that point, so our regular response formula no longer applies and the null and fork description must be used instead. 

The construction is indeed practical.  One first chooses an angular scale that separates large scale image structure from the small scale structure. The image is smoothed on this fixed scale to define $I_{\rm L}$, while the difference between the original and smoothed images defines $I_{\rm S}$. This separation is kept unchanged while the amplitude $A$ of $I_{\rm L}$ is varied. The resulting closure phase change is calculated exactly and compared with the linear prediction in \autoref{eq:large-small-c-res} and the finite response in \autoref{eq:large scale-finite-res}. Repeating the calculation as the two long baselines approach a visibility null then shows how the null increases the sensitivity to changes in the large scale image structure.

It is worth mentioning that for a single static image, $\mathcal R_{\rm L\rightarrow\phi}$ is a deterministic sensitivity: it states how the closure phase would respond if the prescribed broad component changed. It becomes an observable modulation when that component varies between wavelengths, epochs, or members of a source ensemble. The connection to the CMB construction is therefore operational: in both cases, a nonlinear observable is designed to expose information shared between widely separated scales, but the measured quantities and physical settings remain distinct. A numerical realization of this construction using a realistic $M87^*$ data for 2021 is presented in \autoref{fig:cross-scale-cmb-m87}.

Indeed, the squeezed construction above provides an additional route for using the same framework to isolate cross-scale image information. The following examples test the two response hierarchies in controlled source models and in measured interferometric data.


\section{Practical examples}
\label{sec:example1}


\subsection{Closure phase structure near an m-ring null}\label{sec:example-m-ring}

We now apply the fixed image expansion of
\autoref{sec:visib-delta-u} to an m-ring which is an idealized, infinitesimally thin emitting ring of diameter $d$ whose azimuthal brightness variations are represented by a truncated Fourier series \cite{Johnson2020,2022ApJ...930L..15E}. The m-ring provides a useful test of the response formulas because its image structure is simple, while its visibility retains both radial nulls and azimuthal phase information. It also allows the two perturbative viewpoints developed in this paper to be examined separately. We first regard the nonuniform brightness around the ring as a perturbation of a uniform ring. We then hold the complete m-ring fixed and study the closure phase in a
neighborhood of one of its visibility nulls.

\begin{figure}
    \centering
    \includegraphics[width=\linewidth]{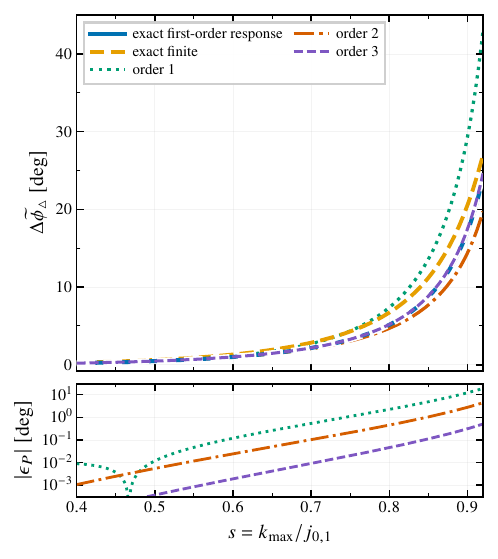}
    \caption{Closure phase response of an m-ring treated as a brightness perturbation of a uniform ring. In the upper panel, the solid curve shows the exact first order response to the complete brightness perturbation, while the long-dashed curve shows the exact finite closure phase change of the full m-ring. The curves labeled order 1, order 2, and order 3 retain Cartesian moments through ranks $N=1$, $N=3$, and $N=5$, respectively. The closure triangles are scaled self-similarly by $s=k_{\max}/j_{0,1}$, where $s=1$ places the longest baseline at the first visibility null of the uniform ring. The lower panel shows the absolute truncation error $|\epsilon_P|=\left| \delta\widetilde{\phi}_{\triangle}^{(P)} \delta\widetilde{\phi}_{\triangle}^{\rm exact,\,1st}\right|$, for the same three moment orders, plotted on a logarithmic scale. Successive odd rank truncations converge rapidly toward the exact first order response; in particular, the order 3 curve nearly overlaps it in the upper panel, while the residual panel makes the remaining difference visible. The separation between the exact first order and exact finite curves measures the nonlinear dependence of closure phase on the brightness perturbation.}
    \label{fig:mring-brightness-moments}
\end{figure}

\subsection{Brightness perturbations of a uniform ring}
\label{sec:m-ring-image-perturbation}

We work in polar coordinates $(r,\varphi)$ on the sky and $(\rho,\psi)$ in the Fourier plane. A thin m-ring of radius $R$, total flux $F$, and azimuthal structure through order $M$ can be written as
\begin{equation}
I_M(r,\varphi)=\frac{F}{2\pi R}\,\delta(r-R)
\left[1+\sum_{m=1}^{M}\left(\beta_m e^{im\varphi}
+\beta_m^\ast e^{-im\varphi}\right)\right].
\label{eq:m-ring-image}
\end{equation}
The coefficients satisfy the reality condition already displayed in \autoref{eq:m-ring-image}. It is useful to write
\begin{equation}
\beta_m=a_m e^{-im\varphi_m},
\end{equation}
so that the brightness modulation at order $m$ is
$2a_m\cos[m(\varphi-\varphi_m)]$. The amplitudes are assumed to be small enough that the total brightness remains nonnegative.

With the Fourier convention used in this paper, the visibility is
\begin{equation}
V_M(\rho,\psi)=F\left[J_0(k)+\sum_{m=1}^{M}i^m J_m(k)\left(\beta_m e^{im\psi}+\beta_m^\ast e^{im\psi}\right)\right],
\label{eq:m-ring-visibility}
\end{equation}
where $k:=2\pi R\rho$, and $J_m$ is the Bessel functions of the first kind. The $m=0$ term is the visibility of a uniform ring, while the remaining terms contain the azimuthal brightness structure\footnote{A circular Gaussian broadening of the ring multiplies \autoref{eq:m-ring-visibility} by a real positive function of $\rho$. It therefore changes the visibility amplitudes but not their phases as we saw in \autoref{sec:visibility-space-invariances}. Therefore, for the closure phase calculation is sufficient to work with the thin ring.}.

We now take the uniform ring as the reference image,
\begin{equation}
I_0(r)=\frac{F}{2\pi R}\delta(r-R),\qquad
V_a^0=FJ_0(k_a),
\label{eq:uniform-ring-ref}
\end{equation}
where $k_a=2\pi R\rho_a$ for the $a$-th baseline. The azimuthal modes define the brightness perturbation $\delta I=I_M-I_0$. Since every $m\neq0$ mode integrates to zero around the ring, this perturbation conserves total flux $M^{(0)}[\delta I]=0$.

At a fixed closure triangle, the fractional visibility perturbation on baseline $a$, away from the radial nulls of the reference ring, $J_0(k_a)\neq 0$, is
\begin{equation}
r_a:=\frac{\delta V_a}{V_a^0}=\sum_{m=1}^{M}i^m
\frac{J_m(k_a)}{J_0(k_a)}\left(\beta_me^{im\psi_a}
+\beta_m^\ast e^{-im\psi_a}\right).
\label{eq:m-ring-fractional-visib}
\end{equation}
The finite change of closure phase relative to the uniform ring is given by \autoref{eq:finite-dlogV}. The reference closure phase is $0$ or $\pi$, depending on the sign of $\prod_aJ_0(k_a)$. For the comparisons below we choose a zero free triangle for which this product is positive, so that the reference closure phase is zero.

For weak azimuthal structure, the first order response follows directly by \autoref{eq:cp-dV-over-V}. Using $\beta_m=a_m e^{-im\varphi_m}$, we obtain
\begin{equation}
\delta\widetilde\phi_\triangle=2\sum_{m=1}^{M}a_m\sin\,\left(\frac{m\pi}{2}\right)\sum_{a=1}^{3}\frac{J_m(k_a)}{J_0(k_a)}\cos\,\left[m(\psi_a-\varphi_m)\right].
\label{eq:m-ring-linear-cp}
\end{equation}
The leading dipolar mode is given by $m=1$ term. An $m=2$ term is absent because a ring that contains only even azimuthal modes is invariant under $\boldsymbol{x} \to -\boldsymbol{x}$. Its visibility is real and its closure phase remains $0$ or $\pi$. An even mode can modify closure phase only through nonlinear coupling(!) with odd modes. 

The same result can be obtained through the Cartesian moment hierarchy. Since for uniform ring all reference visibilities are real, the response tensor \autoref{eq:moment-kernel-coefficients} reduce to 

\begin{equation}
\mathcal K_{\triangle,I_0} =\frac{\sin(N\pi/2)}{N! \,\,F}\sum_{a=1}^3  \frac{\boldsymbol{u_a^{\otimes N}}}{J_0(k_a)}.\label{eq:uni-ring-moment-ker}  
\end{equation}
all even ranks vanish $\mathcal K_{\triangle,I_0}^{(2N)}=0.$ We see that the azimuthal and Cartesian hierarchies are consistent. An azimuthal mode $m$ first appears in Cartesian moment of rank $N=|m|$, and then in rank $N=|m|+2, |m|+4, \it{etc}$. For example dipolar mode contributes at $N=1,3,5,\dots$, where we see in \autoref{fig:mring-brightness-moments}. Summing these ranks reconstructs the Bessel function $J_1$.

\subsection{Moment hierarchy in m-ring null}
\label{sec:m-ring-CP}
We now keep the complete ring fixed and move the observing triangle through a small Fourier plane neighborhood. The simplest source that displays the relevant behavior is again a
ring with a dipolar brightness modulation,
\begin{equation}
I(r,\varphi)=\frac{F}{2\pi R}\delta(r-R)\left[1+2a_1\cos(\varphi-\varphi_1)\right].
\end{equation}
Its visibility is
\begin{equation}
V(\rho,\psi)=F\left[J_0(k)+2ia_1J_1(k)\cos(\psi-\varphi_1)\right].
\label{eq:m1-ring-visib}
\end{equation}
A true visibility null requires both the real and imaginary parts of \autoref{eq:m1-ring-visib} to vanish. The nulls are therefore located at
\begin{equation}
k=j_{0,n}, \qquad \psi_\ast=\varphi_1  \pm \frac{\pi}{2},
\label{eq:m1-ring-null-locs}
\end{equation}
where $j_{0,n}$ is the $n$-th positive zero of $J_0$. The uniform ring produced an entire circular null. The dipolar brightness modulation breaks this null circle into two isolated points as we discussed earlier.

Now we consider $k=j_{0,n}+\delta k$, $\psi=\psi_\ast+\delta\psi$. To first order,
\begin{equation}
V\simeq-F\,J_1(j_{0,n})\left(\delta k\pm2ia_1\delta\psi
\right).
\label{eq:m1-ring-null-normal-form}
\end{equation}
The two nulls carry opposite charges. More importantly for the present application, \autoref{eq:m1-ring-null-normal-form} identifies the image information encoded in the
local phase pattern. The ring radius fixes the radial location of the null through $j_{0,n}=2\pi R\rho_\ast$. The dipole orientation fixes its angular location through $\psi_\ast=\varphi_1\pm\pi/2$. The dipole amplitude $a_1$ fixes the relative rate at which the phase changes in the radial and angular directions.

To construct a local closure phase map, we choose a zero free reference triangle near one of these nulls. We move one baseline through a two dimensional displacement $\boldsymbol h=(h_u,h_v)$, hold the second baseline fixed, and move the third one by $-\boldsymbol h$, so that the triangle remains closed.
Then a two dimensional neighborhood of this triangle is generated by
\begin{equation}
\boldsymbol k_1=\boldsymbol k_1^0+\boldsymbol h,\qquad
\boldsymbol k_2=\boldsymbol k_2^0,\qquad
\boldsymbol k_3=\boldsymbol k_3^0-\boldsymbol h ,    
\end{equation}
The opposite displacements of the first and third
baselines preserve closure. The exact change relative to the reference closure phase is
\begin{equation}
\Delta\widetilde\phi_\triangle^{\rm exact}(\boldsymbol h)=\Im\Log\left[\frac{V(\boldsymbol k_1^0+\boldsymbol h)V(\boldsymbol k_3^0-\boldsymbol h)}{V(\boldsymbol k_1^0)V(\boldsymbol k_3^0)}\right].
\label{eq:mring-exact-deltau-cp}
\end{equation}
The expansion is about the zero free reference triangle. The phase can be followed continuously from $\boldsymbol h=0$ along any path that does not cross the marked visibility null. Now to obtain the local approximations, let

\begin{equation}
D_a^{(N)}:=\nabla_{\boldsymbol k}^{\,N}\Log V(\boldsymbol k)\big|_{\boldsymbol k=\boldsymbol k_a^0}.
\end{equation}
The closure phase expansion through order $P$ in the displacement is then
\begin{equation}
\Delta\widetilde \phi_\triangle^{[P]}(\boldsymbol h)
=\Im \sum_{N=1}^{P} \frac{1}{N!}\left[D_1^{(N)}
+(-1)^N D_3^{(N)}\right] : \boldsymbol h^{\otimes N}.
\label{eq:mring-deltau-expansion}
\end{equation}
The factor $(-1)^N$ follows from the displacement $-\boldsymbol{h}$ of the third baseline. The first term gives the local closure phase gradient, the second term the local curvature and the the third one accounts for the leading asymmetry beyond a quadratic surface.

\begin{figure}
      \centering
     \includegraphics[width=\columnwidth]{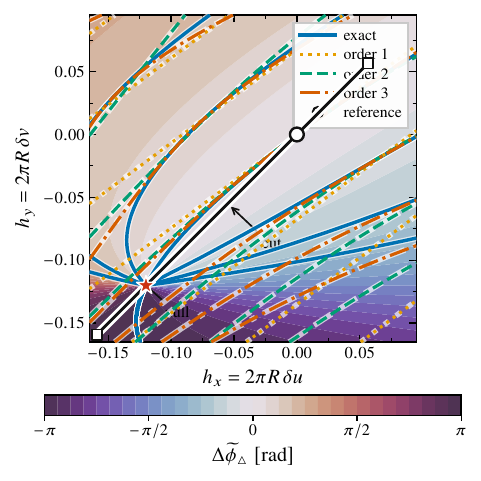}    
      \caption{\label{fig:mring-deltau-map}Closure phase variation for a fixed dipole modulated m-ring in a neighborhood of a visibility null. it shows exact closure phase field in the local $(h_x,h_y)$-plane, together with the first three local approximations. The reference triangle is marked by the filled point, and the nearest visibility null is indicated separately with a star. The solid cut line shows the one-dimensional slice plotted in \autoref{fig:mring-deltau-cut}: it passes through the reference triangle and is directed toward the nearest null. The exact closure phase is undefined at the starred null; the displayed phase field is wrapped to $[-\pi,\pi)$.}
  \end{figure}

\begin{figure}
    \centering
    \includegraphics[width=\columnwidth]{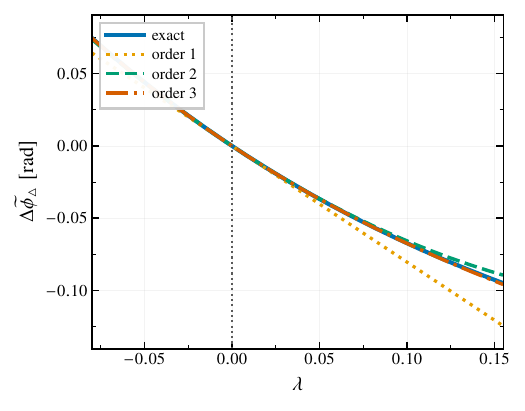}
    \caption{The exact closure phase and its successive local approximations along that cut in \autoref{fig:mring-deltau-map}. Here $\lambda$ is the signed distance along the diagonal cut, measured from the reference triangle toward the nearest visibility null; the plotted interval stops short of the null. Since the expansion is taken about the reference triangle, all approximations agree there, and higher order terms improve the agreement over a larger portion of the cut. The deviation grows as the slice approaches the nearby null, where the regular log-visibility expansion eventually ceases to apply.\label{fig:mring-deltau-cut}}
 \end{figure}

For the numerical comparison, we evaluated \autoref{eq:mring-exact-deltau-cp} and the successive truncations of \autoref{eq:mring-deltau-expansion} on the same grid in $(h_x,h_y)$. The derivatives $D_a^{(N)}$ are obtained directly from the analytic visibility in \autoref{eq:m1-ring-visib}. No image reconstruction or numerical Fourier transform is needed in this example. \autoref{fig:mring-deltau-map} shows the exact closure phase surface together with the approximations through first, second, and third order. The linear approximation reproduces the direction and magnitude of the local phase gradient, but not the bending of the constant phase contours. The quadratic approximation captures this curvature, while the
cubic term recovers most of the remaining asymmetry. For the one dimensional comparison we take a cut through the reference triangle in the direction of the nearest visibility null presented in \autoref{fig:mring-deltau-cut} to show the same convergence more directly. The diagonal cut was chosen because it is the straight line from the reference triangle to the nearest null. Along this direction, the cut approaches the singularity as directly as possible.

Note that the expansion orders in this calculation refer to powers of the Fourier plane displacement $\boldsymbol h$. They are not the azimuthal index of the m-ring and should not be identified with dipole, quadrupole, or higher angular multipoles of the image. The source is fixed throughout; the figures test only how well a local polynomial in baseline position reproduces the exact closure phase near a visibility null.

\medskip

\subsection{Application to the $M87^*$ data}\label{sec:example-m-87}

The m-ring example provided a controlled test in which the image and its visibility were known exactly. We now apply the two response descriptions to the 2019 observations of M87$^\ast$ e.g., \cite{EHT2019IV}.
The first application keeps the source fixed and follows the bispectrum as Earth rotation moves the AA-GL-PV triangle through the Fourier plane. The second application keeps the measured baselines fixed and asks which low rank asymmetric image moments are required to reproduce the observed closure phase. Together, these calculations demonstrate both directions of the theory on realistic interferometric data: the observed complex bispectrum measures the track projected Fourier plane response, while the data constrained image reveals the asymmetric moment structure that builds the closure phase.
\subsubsection{Local motion of the M87* bispectrum}
\label{sec:m87-bispectrum-motion}

The AA-GL-PV triangle shows a rapid closure phase rotation at the same time that one of its visibility amplitudes passes through a deep minimum. This makes it a natural observational test of the local Fourier plane expansion developed in \autoref{sec:visib-delta-u}. No image model is introduced in this calculation. For each scan, we form the complex bispectrum from the three consistently oriented baseline visibilities,
\begin{equation}\label{eq:m87-bis-initial}
B_\triangle(t)=V_{\rm AA-GL}(t)\,V_{\rm GL-PV}(t)\,V_{\rm PV-AA}(t).
\end{equation}
On a time interval over which the bispectrum remains nonzero, its logarithm can be followed continuously:
\begin{equation}
\Log B_\triangle(t)= \ln |B_\triangle(t)| + i \,\widetilde\phi_\triangle(t),
\end{equation}\label{eq:m87-log-bis}
where $\widetilde\phi_\triangle(t)$ is the closure phase followed over selected time interval continuously. Note that $\Log B_\triangle(t)$ allows the bispectrum amplitude and closure phase to be treated as two parts of the same complex response, rather than fitting them independently. Therefore, a local model that reproduces the phase rotation must also reproduce the simultaneous change in amplitude. 
We choose the reference time $t_0$ near the minimum of bispectrum amplitude while  $B_\triangle(t_0)$ remains nonzero, and we define $\tau:=t-t_0$. Along the one dimensional Earth rotation track, the general Fourier plane motion expansion takes this form
\begin{equation}
\Log B_\triangle(t) \simeq C_0+L_\triangle^{[\leq P]},
\label{eq:m87-local-bis-expansion}
\end{equation}
where 
\begin{equation}
L_\triangle^{[\leq P]}= \sum_{n=1}^{P} \frac{C_n}{n!}\,\tau^n,    
\end{equation}
and the coefficients $C_n$ are complex,
\begin{equation}
C_n=\left.\frac{d^n}{dt^n}\Log B_\triangle(t) \right|_{t=t_0},
\label{eq:m87-motion-general-coefficient}
\end{equation}
which are the one dimensional projections of the Fourier plane derivative hierarchy onto the actual Earth rotation track and not independent phenomenological polynomial parameters. Their real and imaginary parts jointly describe the local evolution of bispectrum amplitude and closure phase.
in which their real parts describe the local variation of logarithmic amplitude and imaginary parts describe the unwrapped closure phase
\begin{equation}
|B_\triangle|^{[\leq P]}= \exp\left[\Re (C_0+L_\triangle^{[\leq P]})\right],\quad
\widetilde\phi_\triangle^{[\leq P]}=\Im (C_0+ L_\triangle^{[\leq P]}).
\label{eq:m87-local-logB-observables}
\end{equation}
Thus, each order is fitted once, using a single set of complex coefficients, rather than fitting the amplitude and closure phase independently. The fits are performed separately at each polynomial order. The first order approximation follows the local tangent to $\Log B_\triangle$; the second order approximation adds the bending of the trajectory; and the third order approximation allows the curvature itself to change across the interval. These are the one dimensional projections of the Fourier plane derivative hierarchy along the actual Earth rotation track.

In fact, \autoref{eq:m87-local-bis-expansion} is the one dimensional restriction of the Fourier plane response hierarchy in Equations \eqref{eq:delta-u-visibi}-\eqref{eq:106!} to the measured Earth rotation track; the coefficients $C_n$ are therefore the corresponding track projected visibility derivatives.

\begin{figure*}
    \centering
    \includegraphics[width=0.95\linewidth]{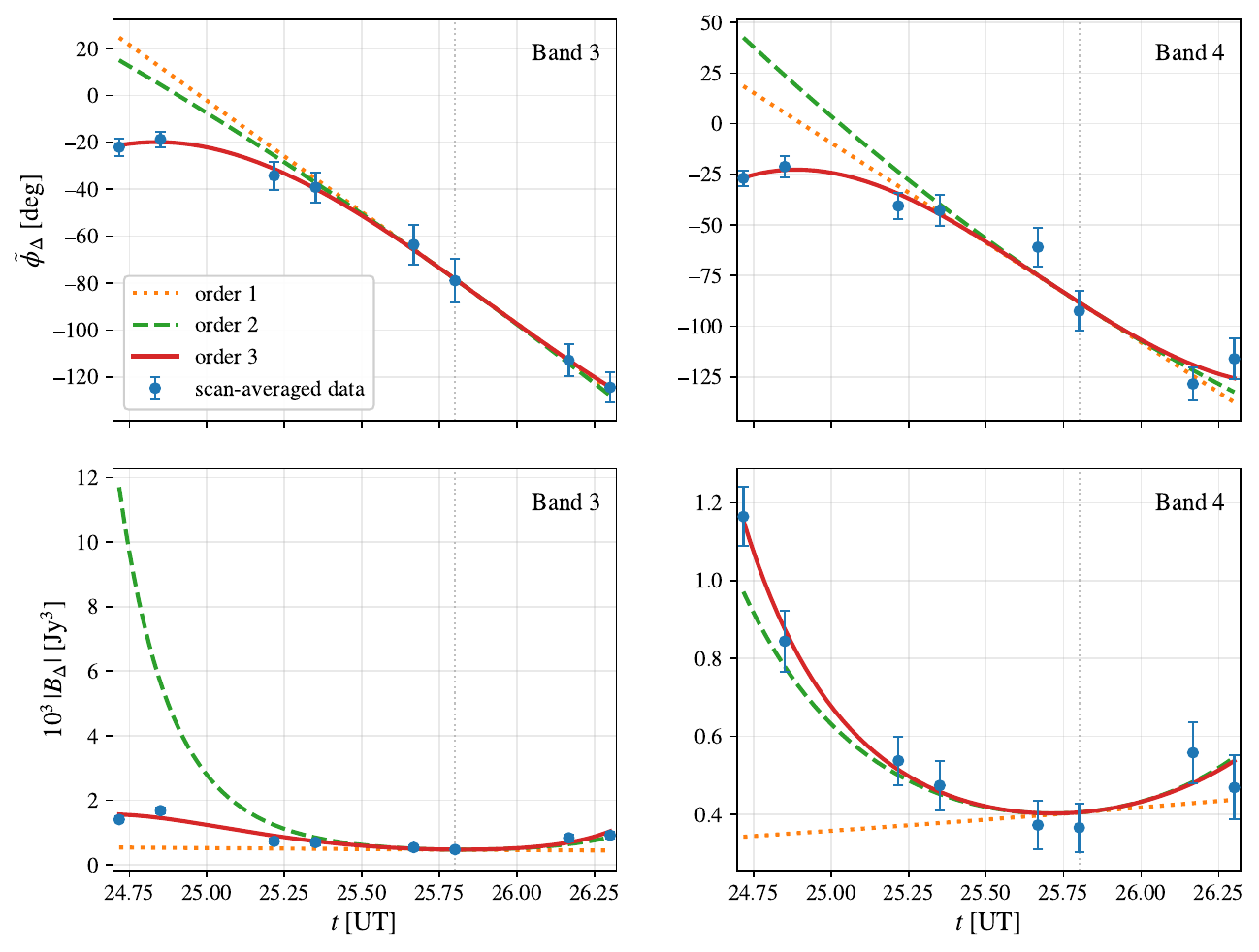}
    \caption{Local Fourier plane expansion of the scan-averaged AA-GL-PV bispectrum in M87* bands 3 and 4. The upper panels show the unwrapped closure phase and the lower panels show the bispectrum amplitude. Points with error bars are the scan-averaged data.
    Dotted, dashed, and solid curves show separate first, second, and third order complex polynomial fits to $\Log B_\triangle(t)$ about $t_0=25.8006$ UT. At each order, the same complex fit generates both the closure phase and bispectrum amplitude curves. The vertical dotted lines mark $t_0=25.8006$ UT. The quadratic term captures the local curvature near $t_0$, while the cubic term extends the agreement across the observed interval, most notably in Band 3. The amplitude and closure phase curves at each order are generated by the real and imaginary parts of the same complex coefficients. A linear variation is insufficient in both bands; the quadratic term recovers the principal curvature, with an additional cubic contribution favored in band 3. The nonzero minimum of $|B_\triangle|$ identifies the feature as a near null passage rather than a
    resolved fork.}
    \label{fig:m87-bispectrum-motion}
\end{figure*}

The comparison is shown in \autoref{fig:m87-bispectrum-motion}. A linear description does not reproduce the observed feature: it misses both the pronounced bending of the closure phase and the recovery of the bispectrum amplitude after its minimum. The quadratic term captures the leading curvature in both bands. The third order term gives the remaining asymmetry of the local trajectory, with its effect most clearly visible in band 3. The
agreement improves in the same way as in the controlled example above, but here the successive orders are determined directly from the observed M87* bispectrum.

The finite minimum of $|B_\triangle|$ is also important. The bispectrum approaches a low amplitude region while its phase rotates rapidly, but it does not reach the origin over the measured interval. The observed feature is therefore the regular, near null counterpart of a closure phase fork. It shows how the fork structure is approached in real data when the observing track passes close to, rather than exactly through, a visibility zero.

\subsubsection{Cross-triangle evolusion on a Themis reconstruction from asymmetric image moments}
\label{sec:m87-deltaI-moments}


As a complementary test we apply the asymmetric image hierarchy to an already available image reconstruction. Purpose of this is to ask how the closure phase behavior of an M87* image is organized by the first few asymmetric moment orders.

We use the best fit {\sc Themis} image of M87* data for 2019 as the full brightness distribution $I(\boldsymbol{x})$. With respect to the center the image is decomposed into its inversion even and odd parts,
\begin{equation}
I(\boldsymbol{x})=I_0(\boldsymbol{x}) + \delta I(\boldsymbol{x}),    
\end{equation}
where $I_0(\boldsymbol{x})= I_0(-\boldsymbol{x})$ and $\delta I(-\boldsymbol{x})=-\delta I(\boldsymbol{x})$. We use different closure triangles. The AA-GL-PV triangle passes close to a visibility minimum and is therefore used to test the local Fourier plane motion expansion. For the image moment hierarchy we use GL-NN-PV, which shows a large closure phase excursion while remaining in a regime where successive Cartesian moment ranks provide a controlled approximation to the exact image visibility. This separation allows each hierarchy to be tested in the regime for which it is designed. We compare that exact image based closure phase with successive truncations of the asymmetric moment hierarchy. 

\begin{figure}
    \centering
    \includegraphics[width=\linewidth]{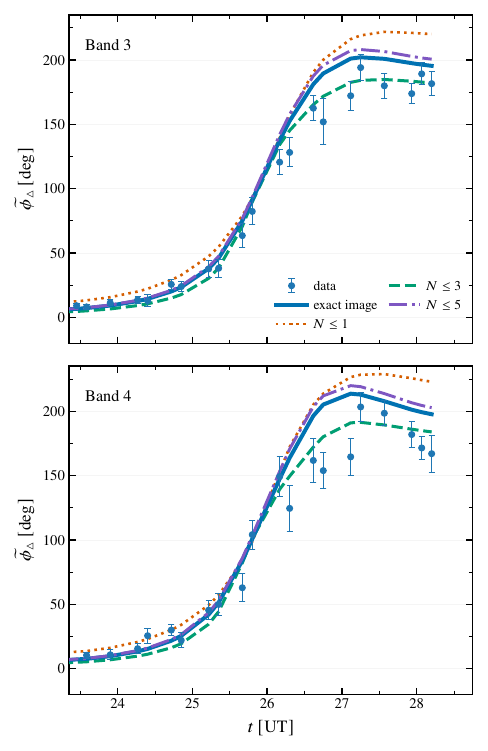}
    \caption{Closure phase on the GL-NN-PV triangle evaluated from the best fit Themis image and from successive Cartesian moment approximations to its inversion odd component. The upper and lower panels show bands 3 and 4, respectively. Points with error bars are the measured scan-averaged closure phases, and the solid blue curve is obtained by Fourier transforming the complete Themis image at the measured baseline coordinates. The remaining curves retain odd Cartesian moments through ranks $N\leq1$, $N\leq3$, and $N\leq5$. The same image moments are used for every scan and in both bands. The exact image reproduces the observed phase rotation in both bands.}
    \label{fig:Themis}
\end{figure}

In \autoref{fig:Themis}
the orange dotted curve retains only the rank one odd moment, the green dashed curve retains odd moments through rank three, and the violet dotted-dashed curve retains odd moments through rank five. In each case, the truncated odd contribution is added to the exact centrosymmetric reference visibility before forming the closure phase. The rank five
approximation follows the exact image prediction to within a few degrees across the
observed track, demonstrating that the asymmetric image structure can be compressed into a finite low rank moment description on this triangle.

Note that the calculation demonstrates that the nonzero closure phase of a realistic resolved image can be compressed into a small number of asymmetric Cartesian moment ranks while retaining the exact centrosymmetric visibility contribution, and should be interpreted as an image to observable consistency test that shows directly how the closure phase signal of a realistic M87* image is built up from the low order asymmetric structure isolated by the hierarchy. 
In this way, the hierarchy is not only an abstract expansion, but also provides a concrete language for describing which parts of a realistic image are responsible for the observed closure phase behavior, even how much of that behavior is already contained in the lowest ranks.

\medskip

\subsection{Beyond mm-VLBI: Phase Closure Nulling}
\label{sec:example-PCN}

In optical and infrared long baseline interferometry, Phase Closure Nulling (PCN) uses a triangle for which one baseline approaches a visibility null of a bright, nearly centrosymmetric source. The correlated flux of the dominant component is then strongly suppressed, while the visibility of a faint asymmetric component remains finite. The closure phase consequently becomes much more sensitive to companions, hotspots, and other weak structures than it is on an unresolved part of the visibility function. This is the central idea of PCN and has been demonstrated observationally \cite{Chelli2009,Duvert2010SPIE,Duvert2010AA}.

The response theory developed above gives a useful way to understand and extend this construction. It shows that PCN is not restricted to a binary model. Near a null, the closure phase amplifies a well-defined projection of an arbitrary brightness perturbation, and the moment hierarchy identifies which parts of that perturbation are being measured.

\medskip

\subsubsection{Null amplification of a faint asymmetric structure}
\label{sec:PCN-kernel}

Let $I_\star(\boldsymbol x)$ show the dominant source and let $\delta I(\boldsymbol x)$ contain the faint surrounding structure. We choose the image origin at the center of the dominant component and assume that $I_\star$ is centrosymmetric. Its visibility is then real,
\begin{equation}
V_a^\star:=V_{I_\star}(\boldsymbol u_a)\in\mathbb R,
\end{equation}
while the total visibility on the $a$-th baseline is
\begin{equation}
V_a=V_a^\star+\delta V_a.
\end{equation}
The finite closure phase of the complete scene is
\begin{equation}
\widetilde\phi_\triangle=\Arg\prod_{a=1}^{3}
\left(V_a^\star+\delta V_a\right).
\label{eq:PCN-finite-general}
\end{equation}
This expression remains meaningful even when a visibility of the primary alone vanishes, provided that the faint component keeps the total visibility nonzero.

Away from the exact primary null, and while $|\delta V_a|\ll |V_a^\star|$, the general response derived above reduces to
\begin{equation}
\delta\widetilde\phi_\triangle=\Im\sum_{a=1}^{3}
\frac{\delta V_a}{V_a^\star}=\int_{\mathbb R^2}
\mathscr K_\triangle^\star(\boldsymbol x)\,
\delta I(\boldsymbol x)\,d^2\boldsymbol x,
\label{eq:PCN-linear-kernel}
\end{equation}
where the kernel takes the particularly simple form
\begin{equation}
\mathscr K_\triangle^\star(\boldsymbol x)
=\sum_{a=1}^{3}\frac{
\sin\,\left(2\pi\boldsymbol u_a\cdot\boldsymbol x\right)}{
V_a^\star}.
\label{eq:PCN-real-kernel}
\end{equation}
\autoref{eq:PCN-real-kernel} makes the PCN amplification explicit. If one baseline approaches a null of the dominant source, its factor $1/V_a^\star$ becomes large and that baseline fringe dominates the response. The closure phase then measures a brightness-weighted sine projection of the faint scene. The amplification is physical, but the apparent divergence of the linear formula is not: once $|\delta V_a|$ becomes comparable to $|V_a^\star|$, the finite expression \autoref{eq:PCN-finite-general} must be used.

Because the reference visibilities are real, the first order closure phase selects only the odd Cartesian moments of the perturbation. Taking the moments about the center of the primary \autoref{eq:perturbation-moments} the hierarchy becomes
\begin{equation}
\delta\widetilde\phi_\triangle=\sum_{\ell=0}^{\infty}\frac{(-1)^\ell(2\pi)^{2\ell+1}}{(2\ell+1)!}
\left[\sum_{a=1}^{3}\frac{\boldsymbol u_a^{\otimes(2\ell+1)}}{V_a^\star}\right]:M^{(2\ell+1)}[\delta I].
\label{eq:PCN-odd-moment-hierarchy}
\end{equation}
Thus PCN does more than indicate the presence of asymmetry. Measurements on several triangles and spectral channels can be organized by Cartesian rank. A rank one model tests whether the signal is described by a simple lopsided component, while rank three and higher ranks test for spatially extended or more structured asymmetry. The nulling baseline provides the sensitivity, and the moment hierarchy provides the image interpretation.

\subsubsection{The unresolved companion limit}
\label{sec:PCN-companion}
The usual binary form of PCN follows immediately as a special case. Let a companion of flux ratio $\epsilon\ll1$ lie at angular separation $\boldsymbol s$ from the center of the primary. Its perturbation is
\begin{equation}
\delta I(\boldsymbol x)=\epsilon F_\star\,\delta^{(2)}(\boldsymbol x-\boldsymbol s),
\end{equation}
and therefore $\delta V_a=\epsilon F_\star e^{\,i\alpha_a}$, $\alpha_a =2\pi\boldsymbol u_a\cdot\boldsymbol s$. Writing $v_a^\star=V_a^\star/F_\star$, the finite closure phase is
\begin{equation}
\widetilde\phi_\triangle=
\Arg\prod_{a=1}^{3}\left(v_a^\star+\epsilon e^{\,i\alpha_a}\right).
\label{eq:PCN-binary-finite}
\end{equation}
When the companion is weak compared with the primary visibility on all three baselines, this reduces to
\begin{equation}
\delta\widetilde\phi_\triangle \simeq \epsilon\sum_{a=1}^{3}
\frac{\sin\alpha_a}{v_a^\star}.
\label{eq:PCN-binary-linear}
\end{equation}
This is the familiar PCN enhancement of a faint companion near a stellar visibility null. In the present language, it is simply the point source specialization of the master kernel. The companion moments are
\begin{equation}
M^{(N)}[\delta I]=\epsilon F_\star\,\boldsymbol s^{\otimes N},
\end{equation}
so the complete moment tower resums to the phase factor
$e^{2\pi i\boldsymbol u\cdot\boldsymbol s}$.

The local form near a null also gives a direct interpretation of the observed phase turn. Let $\rho$ be the spatial frequency coordinate along the nulling baseline, and suppose the primary has a simple zero at $\rho=\rho_0$. In a narrow interval around the null,
\begin{equation}
v_1^\star(\rho) = g(\rho-\rho_0)+ \mathcal O\,\left((\rho-\rho_0)^2\right), \qquad g\neq0.
\end{equation}
If the companion phase varies slowly over this interval, the total visibility on that baseline is
\begin{equation}
v_1(\rho) \simeq g(\rho-\rho_0) + \epsilon e^{\,i\alpha_0}.
\label{eq:PCN-local-visib}
\end{equation}
The corresponding amplitude and closure phase contribution are
\begin{align}
A_1(\rho)&\simeq \left[\left(g(\rho-\rho_0)+\epsilon\cos\alpha_0
\right)^2+\epsilon^2\sin^2\alpha_0\right]^{1/2},
\label{eq:PCN-local-amplitude}
\end{align}
and
\begin{align}
\widetilde\phi_\triangle(\rho) &\simeq\phi_{\rm sm}(\rho)+\operatorname{atan2}\left[ \epsilon\sin\alpha_0,\, g(\rho-\rho_0)+\epsilon\cos\alpha_0\right].
\label{eq:PCN-local-phase}
\end{align}
Here $\phi_{\rm sm}$ contains the slowly varying contribution from the other two baselines. These equations show what the null is measuring. The companion shifts the location of the minimum and prevents the visibility from reaching zero
\begin{equation}
\rho_{\min}-\rho_0 \simeq -\frac{\epsilon\cos\alpha_0}{g},
\qquad A_{\min} \simeq \epsilon|\sin\alpha_0|.
\label{eq:PCN-minimum-observables}
\end{equation}
If the local slope $g$ of the primary visibility is known from the stellar diameter and limb darkening model, the measured shift and minimum depth give
\begin{equation}
\epsilon \simeq \left[ A_{\min}^2 + g^2(\rho_{\min}-\rho_0)^2 \right]^{1/2}.
\label{eq:PCN-contrast-estimator}
\end{equation}
The direction of the phase rotation fixes the sign of $\sin\alpha_0$, while the shift fixes $\cos\alpha_0$. Together they determine the projected companion phase $\alpha_0=2\pi\boldsymbol u_0\cdot\boldsymbol s$, up to the usual fringe ambiguity.
Additional triangles or null orders remove this ambiguity and recover the two dimensional position.

This also clarifies the relation between PCN and a true closure phase fork. For a generic companion, $\sin\alpha_0\neq0$, so the faint component fills the primary null and the visibility trajectory misses the origin. The resulting closure phase is a rapid but continuous turn. A genuine fork of the complete scene occurs only when the perturbing visibility is aligned so that the total visibility can vanish exactly.

\subsubsection{Spectral motion through the null}
\label{sec:PCN-spectral-motion}

In spectrally dispersed optical interferometry, the null is commonly scanned by changing wavelength rather than by changing the physical baseline. For a fixed projected baseline
$\boldsymbol B_a$, $\boldsymbol u_a(\lambda)= 
\boldsymbol B_a/\lambda$
so every spectral channel samples a nearby closed triangle in the Fourier plane. The $\delta\boldsymbol u$ expansion developed above therefore applies directly to PCN spectra.

There are two sources of chromatic closure phase variation. The sampled Fourier point moves as $1/\lambda$, and the source itself may change with wavelength. On a zero free branch these two effects separate exactly
\begin{equation}\label{eq:PCN-spectral-decomposition}
\frac{d\widetilde\phi_\triangle}{d\ln\lambda}
= - \Im\sum_{a=1}^{3} \boldsymbol u_a\cdot \nabla_{\boldsymbol u}\Log V_a + \int_{\mathbb R^2} \mathscr K_{\triangle,I}(\boldsymbol x)\, \frac{\partial I(\boldsymbol x,\lambda)} {\partial\ln\lambda}
\,d^2\boldsymbol x.
\end{equation}
The first term is the known geometric sweep of the observing triangle through the Fourier plane. The second term is the intrinsic chromatic change of the scene. For a companion whose position is fixed but whose flux ratio varies with wavelength, this second term contains its spectrum. \autoref{eq:PCN-spectral-decomposition} gives a practical route to PCN spectroscopy. The visibility amplitudes of the dominant star determine the null position and local slope. The motion term predicts the geometric phase variation across the spectral channels. The remaining signal can then be fitted either by the companion form
\autoref{eq:PCN-binary-finite} or, without assuming a binary, by the odd moment hierarchy \autoref{eq:PCN-odd-moment-hierarchy}. In this way the same data can test whether the asymmetric signal is consistent with a point companion or requires extended structure.

\renewcommand{\arraystretch}{1.5}
\begin{deluxetable*}{lll}
\tablehead{
\colhead{Result} &
\colhead{Required structure} &
\colhead{Scope}}
\startdata
\parbox[t]{0.20\textwidth}{%
\raggedright
Closure phase invariance and wrap diagnostics}
& \parbox[t]{0.29\textwidth}{%
\raggedright
Closed products with station based phase errors}
& \parbox[t]{0.43\textwidth}{%
\raggedright
Applies to coherent interferometric arrays independently of a particular image model.}
\\
\parbox[t]{0.20\textwidth}{%
\raggedright
Fork criterion and correlated baseline signature} &
\parbox[t]{0.29\textwidth}{%
\raggedright
A zero of one visibility factor, and hence of every bispectrum containing that baseline} &
\parbox[t]{0.43\textwidth}{%
\raggedright
Applies across radio, optical/IR, speckle, and aperture masking interferometry.}
\\
\parbox[t]{0.20\textwidth}{%
\raggedright
Null order, winding charge, and partner structure}& \parbox[t]{0.29\textwidth}{%
\raggedright
A complex visibility field; the partner result additionally uses the Hermitian symmetry of the visibility of a real image}
&\parbox[t]{0.43\textwidth}{%
\raggedright
Describes the local topology and global organization of isolated visibility zeros.}
\\
\parbox[t]{0.20\textwidth}{%
\raggedright
Master image kernel and Cartesian moment hierarchy}&
\parbox[t]{0.29\textwidth}{%
\raggedright
The standard Fourier relation between source brightness and complex visibility, evaluated
on a zero free reference triangle}&
\parbox[t]{0.43\textwidth}{%
\raggedright
Connects closure phase changes quantitatively to arbitrary image perturbations and their
spatial moments.}\\
\parbox[t]{0.20\textwidth}{%
\raggedright
Fourier plane motion hierarchy}
&
\parbox[t]{0.29\textwidth}{%
\raggedright
A fixed image sampled on nearby closed triangles}
&
\parbox[t]{0.43\textwidth}{%
\raggedright
Describes Earth rotation tracks, wavelength dependent sampling, and local motion near a
visibility minimum.}
\\
\parbox[t]{0.20\textwidth}{%
\raggedright
Finite near null response}
&
\parbox[t]{0.29\textwidth}{%
\raggedright
A nonzero total visibility, without requiring the fractional perturbation to remain small}
&
\parbox[t]{0.43\textwidth}{%
\raggedright
Describes the amplified but regular regime preceding a true closure phase fork.}
\\[3em]
\enddata
\caption{The first two entries require only the closed product structure of closure phase and station based phase errors. The image response and moment results additionally assume the usual van Cittert-Zernike relation between source brightness and complex visibility.}
\label{tab:result-scope}
\end{deluxetable*}

Phase Closure Nulling is therefore a particularly sharp realization of the general response theory. The null suppresses the dominant symmetric component, the finite log-visibility response describes the large phase turn without a false divergence, the moment hierarchy identifies the image structure being amplified, and the Fourier plane motion expansion separates spectral scanning from intrinsic source chromaticity. The translation between the radio and optical/infrared conventions used in this paper is summarized in \autoref{tab:result-scope}.

\medskip

It is worth mentioning that the same response framework applies to binary systems even when no visibility null is
present. For a faint companion, the primary defines the reference visibility and the secondary provides a perturbation whose Cartesian moments resum to the usual binary phase factor. The master kernel then gives a companion sensitivity pattern and a linear response in the flux ratio. Once a point binary geometry is assumed, however, the finite binary visibility can be written and fitted directly, so the moment representation is most useful when the dominant source is resolved or structurally complex. The nulling regime considered above is the special case in which this faint companion response is strongly amplified. For multi-epoch observations, separating image evolution from Fourier plane motion can distinguish orbital motion from changes in the baseline geometry. A full application requires no additional theoretical development and can be implemented through a joint fit to the visibility amplitudes, phase observables, and a common orbital model.




\section{Conclusion and summary}\label{sec:summary}

This work develops a unified quantitative analytical framework for the rapidly varying features seen in closure phase time series and turns them into a practical, calibration independent diagnostic for visibility nulls. 


The central result of this work is a unified quantitative response theory of closure phase in its regular and singular regimes. Away from visibility zeros, closure phase is a smooth response of
the three complex visibilities and is most naturally described through their logarithms. At a true visibility null, that regular description necessarily fails because the phase itself is undefined. Keeping these two regimes separate provides a clear interpretation of the rapid variations, discontinuities, and fork like structures encountered in interferometric data.

We first established the diagnostic distinction between removable phase wrapping, pseudo nulls, near null phase rotations, and true closure phase forks. A jump caused only
by the principal branch convention can be removed by continuously following the phase. A true fork instead requires the bispectrum to vanish and produces a nonremovable $\pi$-type obstruction. The local null order describes how rapidly the visibility amplitude vanishes, whereas the winding number measures how its phase circulates around the null. These quantities are related but not identical. Simple isolated nulls carry unit charge, while higher order nulls are nongeneric degeneracies that generally split under perturbation, with the total winding preserved. For a real image, Hermitian symmetry further requires every isolated null at $\boldsymbol u_\ast\neq\boldsymbol 0$ to have an inversion partner at $-\boldsymbol u_\ast$ with the same null order and the opposite winding charge. Charged visibility nulls therefore occur in globally neutral pairs, although each member
is locally topologically nontrivial.
For a real image, charged nulls are
also related to opposite charged partners through the Hermitian symmetry of the visibility.

As we remarked at the end of \autoref{sec:order-winding}, it remains unclear precisely how much information is encoded in the locations and winding charges of the nulls.  The surprising amount of mathematical structure that appears with the nulls, including the aforementioned conservation laws, raises the possibility that the nulls contain topological information about the image itself.  We leave answering this question to future work, however.

On a zero free branch, the log-visibility formulation gives a common response law for any change in the three sampled visibilities. This representation also makes the exact
closure phase invariances transparent and separates the visibility response from the physical mechanism that produced it. When the baselines are fixed and the image changes, the response can be pulled back to the brightness distribution. The resulting master kernel gives a spatial map of closure phase leverage: it identifies where a small addition or removal of brightness has the greatest effect on a chosen triangle. Expanding this kernel yields a Cartesian moment hierarchy in which the properties of the triangle and reference image are separated from the spatial moments of the image perturbation. The hierarchy therefore provides a controlled low rank description of which source structures are visible to a given closure phase.

The complementary calculation keeps the image fixed and moves the observing triangle through the Fourier plane. Around a known reference triangle, the three reference
visibilities remain exact, while ordinary image moments describe their variation in a neighboring Fourier plane patch. At every finite order, the displaced visibilities are
explicit polynomials in the reference baselines and their displacements. The corresponding log-visibility expansion gives the local closure phase slope, curvature, and higher order structure along an observing track. This form is directly applicable to Earth rotation synthesis, changes of observing wavelength, and other controlled motions through the Fourier plane. It also shows why the useful neighborhood shrinks near a null: the visibility variation remains regular, but its conversion into phase is amplified by the small reference visibility.

Additionally, squeezed closure triangles provide a direct cross-scale probe: the short baseline supplies the large scale phase reference, while the two long baselines measure the variation of the high spatial frequency phase. The master kernel quantifies how changes in large scale image structure affect this closure phase, with strong enhancement when the long baseline pair approaches a visibility null.

The m-ring examples provide controlled tests of both hierarchies. For brightness perturbations, successive Cartesian moment orders converge toward the exact first order closure phase response, while the remaining difference from the finite result isolates the nonlinear dependence on the perturbation. For motion in the Fourier plane, successive displacement orders recover the local phase gradient, curvature, and higher order asymmetry of the exact closure phase surface near an isolated null. These examples make the two expansions visually and quantitatively distinct: one organizes source structure,
whereas the other organizes local motion of the observing triangle.

The same ideas were then applied to M87* data. The local bispectrum analysis shows that the observed near null phase rotation cannot be described as a purely linear passage through the Fourier plane; curvature is required, with additional higher order structure visible in part of the data. The image perturbation hierarchy provides the complementary source interpretation. Moments inferred from other closure triangles predict the strong AA-GL-PV phase rotation, and the comparison shows that the lowest-rank asymmetric component alone is insufficient. Within the analysis presented here, the principal
feature is captured once the full rank-three Cartesian structure is included, while a further rank five extension gives no clear additional improvement.

Finally, the same framework applies beyond mm-VLBI. Phase Closure Nulling in optical and infrared interferometry is a direct realization of the null-amplified response: a visibility minimum suppresses the dominant nearly symmetric source and enhances the closure phase signature of a faint companion or extended asymmetric component. The master
kernel generalizes this idea beyond binary models, the moment hierarchy organizes the structure being amplified, and the Fourier plane motion expansion separates geometric
scanning of the null from intrinsic spectral changes in the source.


The practical conclusion is that closure phases encode in an immediately accessible form detailed information about the global image structure. The closure phases may be determined jointly by the visibility amplitudes, the station triangles, the motion of the array through the Fourier plane, and their sensitivity to the source structure.  Due to this complexity, closure phases are relegated either to vague indicators of source asymmetry or inputs to detailed image reconstruction algorithms. However, the response theory applies throughout the zero free regime, while visibility minima provide especially sensitive locations where faint or asymmetric structure is strongly amplified. Thus, closure phase can be used not only as an input to image reconstruction or a qualitative asymmetry flag, but as an independent quantitative probe of how source structure and observing geometry combine to produce the measured signal.

\begin{acknowledgments}
This work was supported in part by Perimeter Institute for Theoretical Physics and the University of Waterloo.  Research at Perimeter Institute is supported by the Government of Canada through the Department of Innovation, Science and Economic Development Canada and by the Province of Ontario through the Ministry of Economic Development, Job Creation and Trade.
 A.E.B. receives additional financial support from the Natural Sciences and Engineering Research Council of Canada through a Discovery Grant. 
 \end{acknowledgments}


\appendix 

\section{Convergence of series}\label{app:conve}

If $\delta I$ is supported in a bounded set $\Omega$, and define 
\begin{equation}
\Theta:=\sup_{\boldsymbol x\in\Omega}\|\boldsymbol x\|,
\quad U:=\max_{a=1,2,3}\|\boldsymbol u_a\|,
\quad C_\triangle:=\sum_{a=1}^{3}\frac{1}{|V_a^0|}.
\end{equation}
Then we have
\begin{equation}
    \left|(2\pi)^N\mathcal K_\triangle^{(N)}:M^{(N)}[\delta I] \right|\le C_\triangle\|\delta I\|_1\frac{(2\pi U\Theta)^N}{N!}
    \label{eq:moment-bound-conv}
\end{equation}
Therefore, the moment expansion is absolutely convergent for $\delta I$ with compact support and with finite baselines. A finite truncation is also accurate when the neglected tail is small which is controlled by the dimensionless scale of $(2\pi U\Theta)$ and also by factor $C_\triangle$ which grows when one of the visibilities becomes small enough. More precisely, if the hierarchy is truncated after order $N$, the remaining tail satisfies

\begin{equation}
\left|
\sum_{n=N+1}^{\infty}(2\pi)^n\mathcal K_\triangle^{(n)}:M^{(n)}[\delta I]
\right|\le C_\triangle\|\delta I\|_1\sum_{n=N+1}^{\infty}
\frac{(2\pi U\Theta)^n}{n!}.
\end{equation}
In particular, with $q:=2\pi U\Theta$,
\begin{equation}
|R_{N+1}| \le C_\triangle\|\delta I\|_1
\frac{q^{N+1}}{(N+1)!}e^q .
\end{equation}

\bibliographystyle{plainnat}
\bibliography{CP-references}

\end{document}